\documentclass[prb,twocolumn,aps,superscriptaddress]{revtex4-2}
\usepackage{comment}
\usepackage{graphicx}
\usepackage{dcolumn}
\usepackage{bm}
\usepackage{amssymb}
\usepackage{amsmath}
\usepackage{physics}
\usepackage{sublabel}
\usepackage[utf8]{inputenc}
\usepackage{hyperref}
\usepackage[usenames, dvipsnames]{color}
\usepackage[english]{babel}
\usepackage[T1]{fontenc}
\usepackage{bbm}

\usepackage{float}
\usepackage{verbatim}
\hypersetup{colorlinks = true, linkcolor=blue, citecolor=blue, urlcolor=blue}

  \makeatletter
    \renewcommand\@make@capt@title[2]{%
     \@ifx@empty\float@link{\@firstofone}{\expandafter\href\expandafter{\float@link}}%
      {\textsc{#1}}\@caption@fignum@sep#2\quad}%
    \makeatother

\begin{document}


\title{Resonant weak-value enhancement for solid-state quantum metrology
}

\author{Mahadevan Subramanian}
\affiliation{Department of Physics, Indian Institute of Technology Bombay, Powai, Mumbai-400076, India}

\author{Amal Mathew}
\thanks{Current Address: Department of Applied Physics, Stanford University, Stanford, CA 94305, USA}
\affiliation{Department of Physics, Indian Institute of Technology Bombay, Powai, Mumbai-400076, India}

\author{Bhaskaran Muralidharan}
\affiliation{Department of Electrical Engineering, Indian Institute of Technology Bombay, Powai, Mumbai-400076, India}
\affiliation{Centre of Excellence in Quantum Information, Computation, Science and Technology, Indian Institute of Technology Bombay, Powai, Mumbai-400076, India}
\email{bm@ee.iitb.ac.in}

\date{\today}
\begin{abstract}
Quantum metrology that employs weak-values can potentially effectuate parameter estimation with an ultra-high sensitivity and has been typically explored across quantum optics setups. Recognizing the importance of sensitive parameter estimation in the solid-state, we propose a spintronic device platform to realize this. The setup estimates a very weak localized Zeeman splitting by exploiting a 
resonant tunneling enhanced magnetoresistance readout. We establish that this paradigm offers nearly optimal performance with a quantum Fisher information enhancement of about $10^4$ times that of single high-transmissivity barriers. The obtained signal also offers a high sensitivity in the presence of dephasing effects typically encountered in the solid state. These results put forth definitive possibilities in harnessing the inherent sensitivity of resonant tunneling for solid-state quantum metrology with potential applications, especially, in the sensitive detection of small induced Zeeman effects in quantum material heterostructures. 
\end{abstract}

\maketitle
\section{Introduction}
\begin{figure*}[t]
    \centering
    \includegraphics[width=0.9\textwidth]{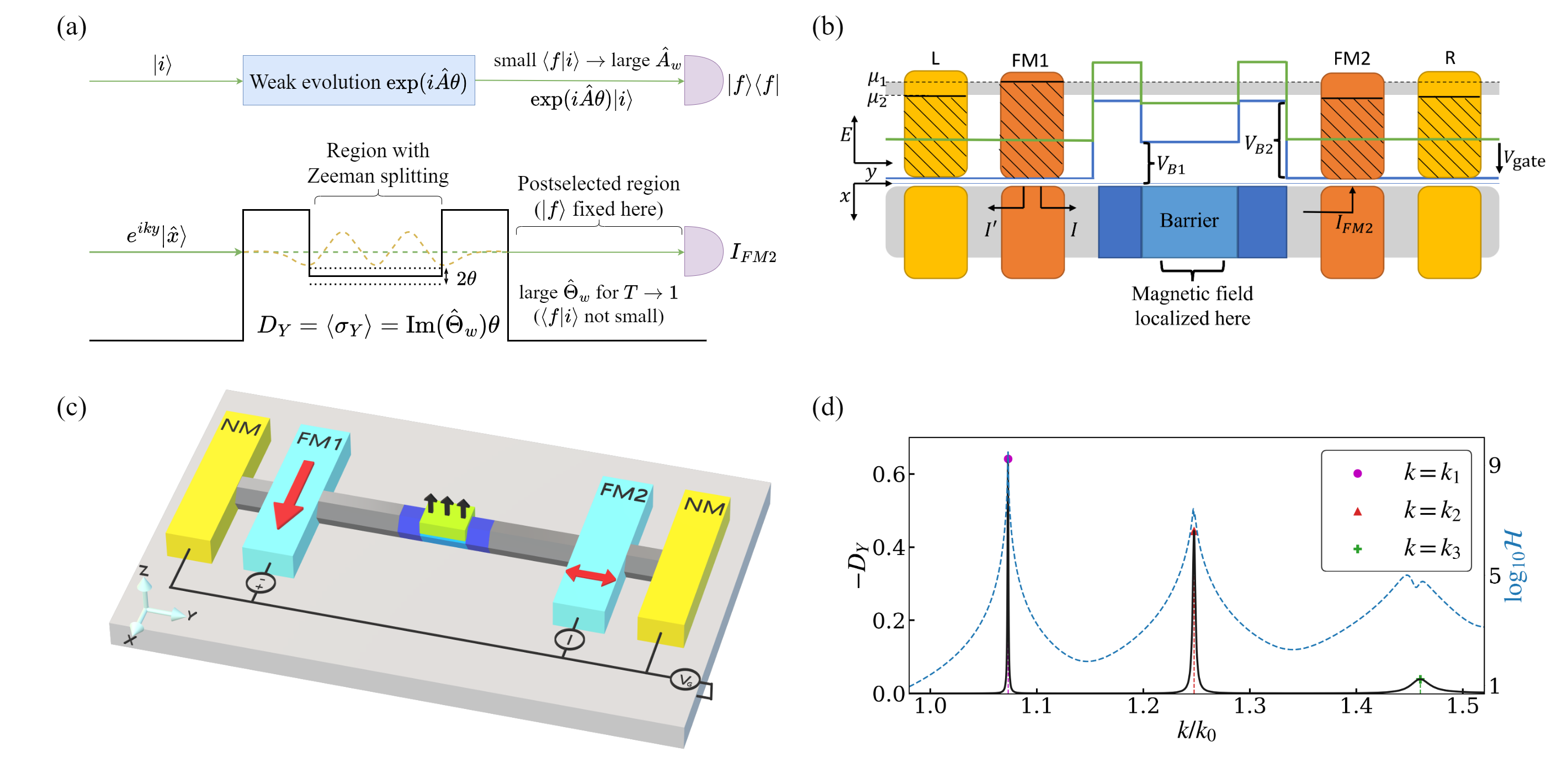}
    \caption{Preliminaries and magnetoresistance signal. (a) A simple schematic (top) representing the weak-value, and the sensing task (bottom) for estimating any localized Zeeman splitting inside the resonant tunneling barrier. 
    General weak-value enhancement techniques involve post-selecting a state $\ket{f}$ 
    Our setup features an enhancement of the weak-value $A_w$ by varying the initial state via a choice of the wave-vector $k$. Contrary to typical setups, the weak-value is enhanced via a choice of $\ket{i}$ although $\langle f|i\rangle$ is not small in general. (b) Detailed device schematic with description of the embedded barrier region. The bottom gate voltage tunes a specific $k$ value via a gate potential $V_\text{gate}$ and a small bias voltage $\mu_1-\mu_0$ selects out the stream outgoing stream. (c) Device schematic for the 1-D channel. There are four contacts, two NM (colored yellow) and two FM (colored red) in direction depicted by the blue arrows. Current readings are obtained from the contact $FM2$. (d) A summit result depicting the signal $D_Y$ as a function of $k$ plotted along with the QFI, shown as $\log_{10}\mathcal{H}$. We notice that there are three characteristic peaks for the $k$ values for which resonant tunneling occurs which are $k_1,k_2$ and $k_3$ that are also values where the QFI takes large values.} \label{fig:Figure1}
\end{figure*}

Quantum metrology 
\cite{PhysRevLett.96.010401,giovannetti2011advances,2017Degen} provides the means toward high-sensitivity parameter estimation using a quantum state as a probe, followed by measurements, and has been demonstrated in a variety of systems \cite{doi:10.1116/5.0007577, TAYLOR20161,PhysRevLett.107.083601,PhysRevA.90.022117,PhysRevX.11.041045,Marciniak_2022}. It is also well established that weak-values can inextricably be linked with quantum sensing \cite{2011Hofmann,2012Kofman,dressel2012weak}. 
The use of weak-values in quantum sensing has typically been explored using quantum optics setups \cite{PhysRevLett.114.170801,PhysRevA.92.032127,PhysRevLett.125.080501,PhysRevA.106.022619}. An important metric to benchmark the quantum sensor performance is the quantum Fisher information (QFI) \cite{Liu_2019,paris2009quantum,FACCHI20104801,FUJIWARA1995119}, which can also be linked to weak-values \cite{2011Hofmann}. The enhancement of weak-values have shown clear experimental advantages for quantum sensing as demonstrated in many works \cite{PhysRevA.91.062107,vaidman2017weak,PhysRevLett.102.173601}, despite theoretical studies which point to how post-selection is disadvantageous, mainly because of a loss in QFI \cite{PhysRevLett.112.040406,combes2014quantum}. This discrepancy has been explored thoroughly with ways to surmount these disadvantages \cite{2014Jordan,KneeCombesFerrieGauger+2016} and methods to increase detection probability as well \cite{vetrivelan2022near}.\\
\indent 
Solid state setups have recently garnered a lot of attention as pivotal testbeds for foundational quantum concepts, such as, quantum state tomography of electrons \cite{jullien2014quantum,samuelsson2006quantum}, entanglement-generation by Cooper pair splitting \cite{PhysRevB.104.245425,ranni2021real,PhysRevB.91.094516,PhysRevLett.127.237701,deacon2015cooper}, and even loophole-free Bell test experiments \cite{pfaff2013demonstration,PhysRevA.63.050101,PhysRevB.83.125304}. 
Given recent advancements in quantum materials and devices, there exist numerous applications that a quantum sensor could provide with its inherent quantum advantage that includes the detection of induced Zeeman splitting in van der Waals heterostructures \cite{zhou2019spin,PhysRevB.102.081403,PhysRevLett.113.266804,PhysRevLett.122.127401,dankert2017electrical,khokhriakov2020gate,kamalakar2016inversion},  the precise estimation of the Rashba spin-orbit coupling parameter \cite{PhysRevB.70.115316,sanchez2013spin}, to name a few. In this work, we demonstrate how double barrier resonant tunneling in the solid-state can be exploited for high-sensitivity detection of localized Zeeman splittings due to an enhanced weak-value, via a magnetoresistance measurement. The setup we propose builds on a generalized four-terminal spin-transport setup \cite{dankert2017electrical,khokhriakov2020gate,kamalakar2016inversion} where the magnetoresistance measurement is directly related to a weak-value $A_w$ \cite{Steinberg1995,PhysRevB.105.144418}, as a measurement outcome of an operator $\hat{A}$ where $\ket{i}$ is an pre-selected state and $\ket{f}$ is the post-selected state:
\begin{equation}\label{eq:weak1}
   A_w = \frac{\langle f|\hat{A}|i\rangle}{\langle f|i\rangle}.
\end{equation}
We now refer to Fig.~\ref{fig:Figure1}(a), which shows how our approach for enhancing weak-values differs from the general approach of post selecting $\ket{f}$ for a small $\langle f|i\rangle$ \cite{combes2014quantum}. By the nature of our setup, the only control we have is changing the incident wave-vector and as it turns out the choice corresponding to the resonant tunneling wave-vectors have the highest weak-value despite having the largest $\langle f|i\rangle$ overlap via close to unity transmission. \\
\indent Our approach provides means to enhance both the weak-values in tandem with increasing sensitivity, via an enhancement in the QFI. We make use of resonant tunneling energy channels \cite{PhysRevB.29.1970} using a double-barrier setup \cite{sun1998resonant,bjork2002nanowire,663544}, thereby allowing Fabry-P\'erot resonances at specific energies. 
The schematic of the double barrier device is described in Fig.~\ref{fig:Figure1} (b) and Fig.~\ref{fig:Figure1} (c).
We also quantify our design with the QFI and further analyze the effects of phase breaking \cite{DANIELEWICZ1984239,datta1997electronic,book91179691,PhysRevB.75.081301,7571106,doi:10.1063/1.5023159,doi:10.1063/1.5044254,PhysRevApplied.8.064014,camsari2020,Duse_2021,PhysRevB.105.144418} that are typically detrimental in such solid-state systems. Our results put forth definitive possibilities in harnessing the inherent sensitivity of resonant tunneling for solid-state quantum metrology with potential applications, especially, in the sensitive detection of small induced Zeeman effects in quantum material heterostructures. 
\section{Setup and Formulation}\label{sec:formulation}
\subsection{The Magnetoresistive setup}\label{subsec:setup}
The device setup schematized in Fig.~\ref{fig:Figure1}(b) and Fig.~\ref{fig:Figure1}(c) consists of a long 1-D nanowire with an embedded barrier region, facilitated electrostatic gating. The embedded region consists of three rectangular barriers with heights $V_{B2}$, $V_{B1}$ and $V_{B2}$ with the total width being $d_2$ and the width of the middle region being $d_1$. The middle region features a magnetic field $B$ along $\hat{z}$, which models for instance a weak Zeeman splitting that is to be estimated precisely, denoted by $V_Z = g\mu_BB$. This multi-terminal setup is a 1-D proof-of-concept which is quite realizable using 1-D nanowires or 2-D structures with multiple gates \cite{dankert2017electrical,khokhriakov2020gate,kamalakar2016inversion} and has been quite intensely pursued \cite{dankert2017electrical,khokhriakov2020gate,kamalakar2016inversion}, especially in situations where induced Zeeman effects occur in localized regions. \\
\indent We can now define the channel Hamiltonian as follows
\begin{align}\label{eq:Hamiltonian}\hat{H} = \begin{cases}\left(\frac{p^2}{2m} + V_{B1}\right)\mathbb{I} - \frac{V_Z}{2}\sigma_z & |y|\leq \frac{d_1}{2}\\\left(\frac{p^2}{2m} + V_{B2}\right)\mathbb{I} & \frac{d_1}{2}<|y|\leq \frac{d_2}{2}\\\left(\frac{p^2}{2m}\right)\mathbb{I} & |y|>\frac{d_2}{2}\end{cases}.\end{align}
The Hamiltonian can be written as $\hat{H} = \hat{H}_0\mathbb{I} + \theta\hat{H}_1\sigma_Z$ where $\hat{H}_0$ and $\hat{H}_1$ are spatial Hamiltonians and $\theta = V_Z/2t_0$. The Zeeman splitting is only in the region where $\hat{H}_1$ is non-zero. As depicted in Fig.~\ref{fig:Figure1}(a), the incident beam of electrons are $+\hat{x}$ spin polarized. The expectation value $\langle\sigma_Y\rangle$ gives us a signal in relation to $\theta$ that depicts the precession of the spin. Our simulations are conducted with the following parameters: 
hopping energy $t_0 = 3.875$ eV, $d_1 = 40$ $nm$ and $d_2 = 80$ $nm$. \\
\indent We use two normal metallic contacts (NM) on the ends of the channel to manipulate reflections in order to make the correct post-selection and the detection of the transport signal feasible \cite{PhysRevB.105.144418}. The ferromagnetic contact $FM1$ introduces $\hat{x}$-polarized electrons facilitated via the bias situation. The current readouts are taken at the ferromagnetic contact $FM2$. The alignment of $FM2$ is along $\pm\hat{y}$. We denote the current readout from $FM2$ in $\pm\hat{y}$ as $I_{FM2}^\pm$. \\
\subsection{Weak-values and sensing}\label{subsec:weakVal}
The estimation task at hand is described in Fig.~\ref{fig:Figure1}(a). Using pre-selection and post-selection of quantum states, one can obtain measurement outcomes outside of the eigenspectrum which can be explained using the concept of weak-values \cite{PhysRevLett.60.1351,PhysRevD.40.2112}. This treatment uses a quantum mechanical pointer which gives the measurement outcomes after being coupled to the system using a von Neumann interaction scheme. The relevance of these results have been discussed with examining how weak measurement cannot be treated as a measurement in a true sense \cite{PhysRevLett.62.2325}.\\
\indent A simpler treatment to weak-values can be found via a perturbative approach \cite{2017Degen,dressel2012weak,Dressel2014}. For an operator $\hat{A}$, the $n^{\mathrm{th}}$ order weak-value is defined to be $A^n_w = \langle f|\hat{A}^n|i\rangle/\langle f|i\rangle$, where $\ket{i}$ is the initial state and the post selection is done with state $\ket{f}$. We define $P = \langle f|i \rangle$ and $P = \langle f|\hat{U}|i\rangle$ where $\hat{U} = \exp(\iota\epsilon\hat{A})$. We can treat $\epsilon$ as a small parameter and perform a Taylor expansion for $U$ and obtain
\begin{equation}\label{eq:perturbWeak}
    \frac{P_\epsilon}{P} = 1 + 2\epsilon\Im A_w - \epsilon^2[\Re A_w^2 - |A_w|^2] +\mathcal{O}(\epsilon^3).
\end{equation}
\begin{figure*}[ht]
    \centering
    \includegraphics[width=0.9\textwidth]{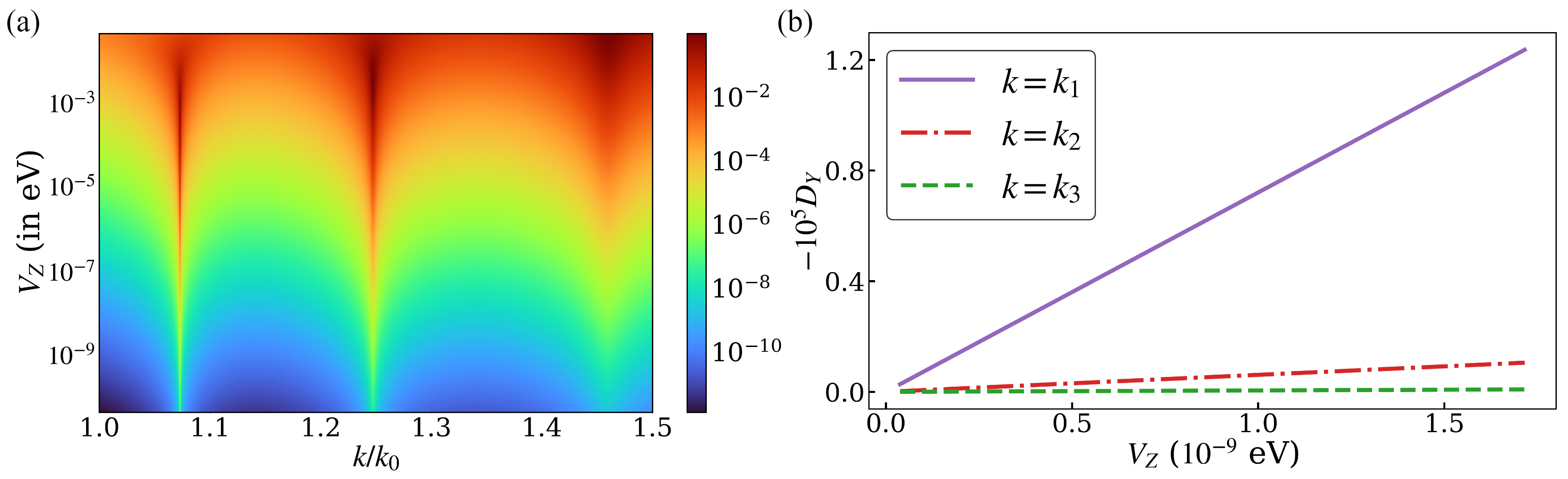}
    \caption{Resonant enhancement in the transport signal. (a) Contour plot depicts the dependence of the signal $D_Y$ with respect to the Zeeman splitting energy $V_Z$. The signal is significantly amplified at the resonant tunnelling wave-vectors as can be seen by the two sharp peaks in the contour. (b) The values of the signal at the resonant values shows a very large amplification for small values of the $V_Z$.} \label{fig:Figure2}
\end{figure*}
To ensure the validity of the weak interaction regime, the quantity $2\epsilon\Im A_w$ must be much larger in magnitude than the sum of all the higher order corrections that follow, which puts a limit to increasing the sensitivity using weak-values \cite{PhysRevD.40.2112}.\\
\subsection{Transport formulation}\label{subsec:transport}
To model the terminal current readout at $FM2$, we employ the Keldysh non-equilibrium Green's function (NEGF) technique \cite{datta1997electronic,Meir-Wingreen-1992,book91179691,haug2007quantum}, whose specific implementation for related setups is elaborated in the Appendix of Ref. \cite{PhysRevB.105.144418}. We go over the brief procedure as follows. The electron correlator is defined as $\mathbf{G}^n = -i \mathbf{G}^< =\mathbf{G}^r\boldsymbol{\Sigma}^{in}\mathbf{G}^a$, where $\mathbf{G}^<$ is the {\it{lesser}} Green's function. Here, the retarded Green's function, $\mathbf{G}^r = [E - \hat{H} - \boldsymbol{\Sigma}]^{-1}$, where $\hat{H}$ is the channel Hamiltonian, $\boldsymbol{\Sigma}$ is the sum of all self-energies, and $\boldsymbol{\Sigma}^{in}$ is the in-scattering function. The quantity $\mathbf{G}^a$ is the hermitian conjugate of $\mathbf{G}^r$  \cite{camsari2020}. The terminal currents are then defined as $I^\pm_{FM2} = \text{Tr}(\Gamma^\pm_{FM2}\mathbf{G}^n)$. For a $\pm\hat{y}$-polarized contact, the expression for the broadening function $\Gamma^\pm_{FM2}$ is a matrix that is only non-zero in the submatrix for the position of the $FM2$ contact on the channel where it takes on value $-t_0e^{ika}(\mathbb{I}+\sigma_Y)/2$. Given that $\rho = \mathbf{G}^n/\text{Tr}(\mathbf{G}^n)$, current measurements of $I^\pm_{FM2}$ are proportional to the probabilities for $\pm\hat{y}$-polarization at the position of the $FM2$ contact as is apparent from the form of its expression. \\
\indent We now define our primary magnetoresistance signal, $D_Y$, which is obtained out of the current readouts from the contact $FM2$ when it is $\pm y$ polarized and defined as
\begin{equation}
    D_Y = \frac{I^+_{FM2} - I^-_{FM2}}{I^+_{FM2} + I^-_{FM2}}.
\end{equation}
 From our physical understanding of the current measurements, the signal $D_Y$ is proportional to the average value $\langle\sigma_Y\rangle$. Let $\ket{\psi}$ be an eigenstate of $\hat{H}_0$ with an energy $\varepsilon(k)$ and $\ket{\psi^\pm}$ is the scattered wavefunction obtained for Hamiltonian $\hat{H}_0\pm\hat{H}_1$. The scattered waves can be calculated using the equilibrium Green's function $\hat{G}_0$ evaluated from $\hat{H}_0$ \cite{e22111321,sakurai_napolitano_2017}. We define $\ket{f}$ as the momentum eigenstate with wave-vector $k$ multiplied by a Heaviside-step function to make it zero everywhere except to the right of the barrier. By taking the Born approximation, the first order approximation for $D_Y$ is as follows (see Appendix \ref{app:scatter} for a more detailed discussion) :
\begin{equation}\label{eq:weakDy}
    D_Y = -\frac{V_Z}{2t_0}\Im\left(\dfrac{\langle f|\hat{G}_0\hat{H}_1|\psi\rangle}{\langle f|\psi\rangle}\right) + \mathcal{O}(\theta^2).
\end{equation}
This elucidates that amplifying the imaginary part of the weak-value for $\hat{G}_0\hat{H}_1$ can boost the sensitivity of $D_Y$ with respect to $\theta$. This weak-value also has physical relevance as a form of the tunneling time as explored in \cite{Steinberg1995}. It has been established that $D_Y = -\omega_L\tau_Y$ where $\tau_Y$ is a real part of the weak-value of the barrier potential \cite{Steinberg1995,PhysRevB.105.144418} which can be proven to be equivalent to \eqref{eq:weakDy}. This notion can be generalized in the case of more complicated barriers which would only change the Green's function $\hat{G}_0$, while $\hat{H}_1$ takes into account the localized Zeeman splitting.
\subsection{Quantum Fisher information}\label{subsec:QFI}
The task of quantum sensing is fundamentally a parameter estimation task and the QFI is a very relevant figure-of-merit \cite{Liu_2019,paris2009quantum}. In a general estimation task, a set of measurements are performed on a parameterized state to retrieve information on the parameters. We focus on the single parameter case, relevant to our setup. The symmetric logarithm derivative \cite{FUJIWARA1995119, Liu_2016}, denoted as $L_\theta$, for the estimation task for a parameterized state $\rho_\theta$ is defined by the equation $\partial_\theta\rho_\theta = \frac{1}{2}(L_\theta\rho_\theta + \rho_\theta L_\theta)$. The QFI denoted by $\mathcal{H}$, is defined as $\mathcal{H} = \text{Tr}(L_\theta^2\rho_\theta)$, where \indent $\mathcal{H}$ will always be bounded above by the maximum eigenvalue of the operator $L_\theta^2$. \\
\indent Given $\partial_\theta\rho_\theta$ and $\rho_\theta$, we can find $L_\theta$ as a solution to a continuous Lyapunov equation \cite{Liu_2016}. As established in the previous section, we can write the density matrix $\rho = \mathbf{G}^n/\text{Tr}(\mathbf{G}^n)$. We define the parameter to estimate as $\theta$ where $V_Z = \theta t_0$. From this, we can use the NEGF equations to obtain the expressions
\begin{gather}
\tilde{L} = H_1\mathbf{G}^r - \frac{\text{Tr}H_1\mathbf{G}^r\mathbf{G}^n)}{\text{Tr}(\mathbf{G}^n)}\mathbb{I},\\
    \partial_\theta\rho = \tilde{L}\rho + \rho\tilde{L}^\dagger.
\end{gather} 
The classical Fisher information (CFI) \cite{FACCHI20104801} for this parametrized state can also be obtained by using the current measurements $I_{FM2}^\pm$ to define a classical probability distribution since currents at the $FM2$ contact behave like a positive operator-valued measure (POVM) for measurements along $+\hat{y}$ and $-\hat{y}$ (see discussion in Appendix \ref{app:measure}). Since we obtain current measurements from the contact, they will be in ratio of the probabilities obtained from this POVM, which can be used for ascertaining the CFI. More discussions on obtaining the CFI and QFI can be found in Appendix \ref{app:QFI} and Appendix \ref{app:CFI}.\\
\indent We denote the CFI as $\mathcal{H}_c$ which is dependent on the POVM set that is chosen. The QFI can be equivalently defined as the maximal CFI over all POVMs, hence $\mathcal{H}_c\leq\mathcal{H}$ \cite{Liu_2019}. The quantum Cram\'er-Rao bound \cite{PhysRevLett.98.090401,BRAUNSTEIN1996135} gives us a minimum bound on $\Delta\hat{\theta}$ where $\hat{\theta}$ is an unbiased estimator for $\theta$ for $M$ repetitions of the measurements. Picking a better POVM will result in a better $\mathcal{H}_c$, which gives a better bound on $\Delta\hat{\theta}$, as seen by the inequality
\begin{equation}\label{eq:cramer_rao}
    (\Delta\theta)^2 \geq \frac{1}{M\mathcal{H}_c}\geq \frac{1}{M\mathcal{H}}.
\end{equation}
\indent Another common metric for the performance is that of the signal to noise ratio \cite{agarwal2022quantifying}. This is linked to the QFI using a measure defined as $R_\theta = \theta^2/\Delta\theta$. From \eqref{eq:cramer_rao}, we get $R_\theta = \leq \theta^2\mathcal{H}$. The quantity $\theta^2\mathcal{H}$ is also referred to as the estimability of the parameter \cite{paris2009quantum}. Our setup has practically unlimited repeated measurements since we obtain steady state current measurements. Since our signal is proportional to our parameter and the measurements are uncorrelated, $R_\theta$ would scale linearly with $N$ with $N$ probes. In what is known as the Heisenberg limit, the scaling of $R_\theta$ goes as $N^2$ which is not possible here since that would require correlations between the probes \cite{demkowicz2012elusive, PhysRevA.85.042112, PhysRevLett.105.180402}. 

\begin{figure}[h]
    \centering
    \includegraphics[width=\linewidth]{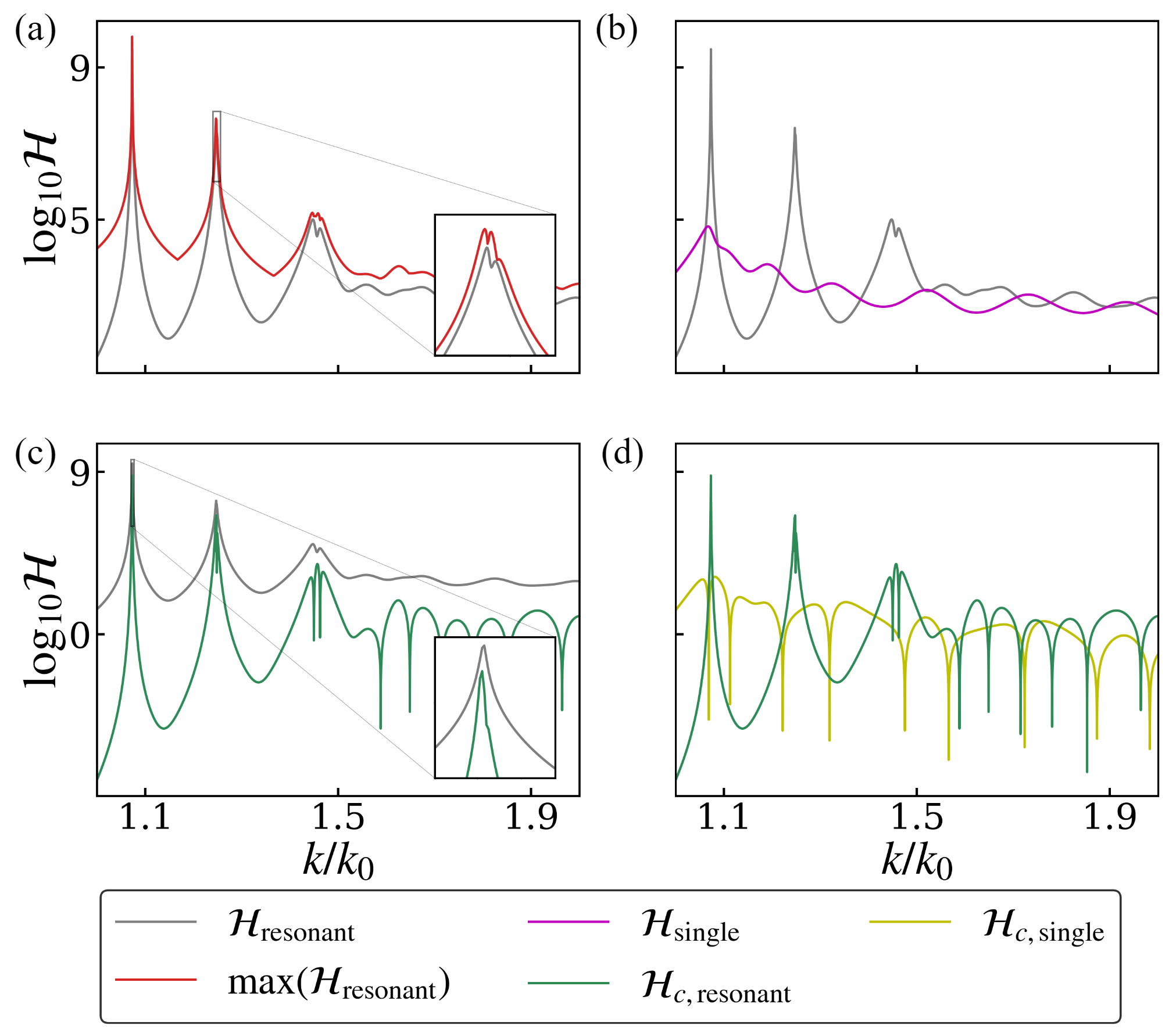}
    \caption{QFI and CFI of the parametrized state in the setup. (a) Comparison between the QFI of the resonant tunneling setup (labelled $\mathcal{H}_\mathrm{resonant}$) to the maximum possible QFI for the same setup (labelled $\max(\mathcal{H}_\mathrm{resonant}$)). This demonstrates that the QFI approaches closely the limiting value close to a resonant tunneling wave-vector, say, $k_2$ (see inset). (b) Comparison of QFI for resonant tunneling setup to the QFI for the single barrier setup (labelled $\mathcal{H}_\mathrm{single}$). It can be seen that the resonant tunneling setup clearly outperforms the single barrier setup. (c) Comparison between the CFI (labelled $\mathcal{H}_{c,\mathrm{resonant}}$) and the QFI of the resonant tunneling setup. We notice that the CFI almost approaches the QFI for resonant tunneling wave-vectors (see inset for $k_1$). (d) Comparison between the CFI of the resonant tunneling setup and the single barrier setup $\mathcal{H}_{c,\mathrm{single}}$, again demonstrating that the resonant tunneling setup outperforms the single barrier setup even here.} \label{fig:Figure3}
\end{figure}
\begin{figure}[h]
    \centering
    \includegraphics[width=\linewidth]{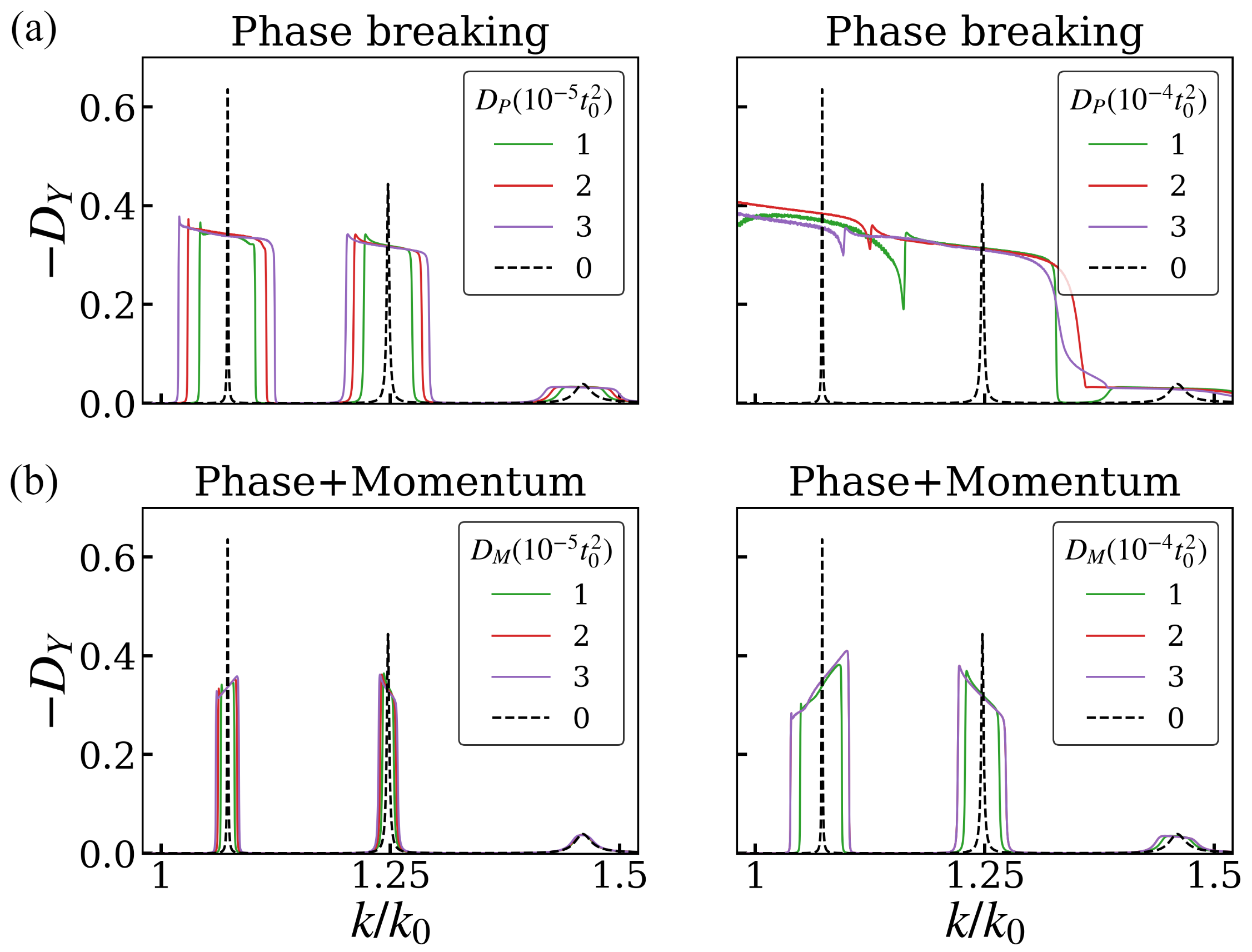}
    \caption{Effects of phase relaxation and momentum relaxation.  (a) Results for the resonant tunnelling setup with non-zero values of $D_P$ which causes pure-phase dephasing, and (b) the results with non-zero values of $D_M$ which cause momentum + phase relaxation} \label{fig:Figure4}
\end{figure}
\section{Results}\label{sec:results}
\subsection{Response of the sensor and the quantum Fisher information}\label{subsec:response}
The signal $-D_Y$ obtained for a Zeeman splitting of $V_Z = t_0/5000$ is depicted in Fig.~\ref{fig:Figure1}(d) and this shows us three values of the wave-vector $k$ where the signal is very clearly amplified. 
The wave-vector $k_3$ has a higher energy than $V_{B2}$ which does not correspond to resonant tunneling. Additionally, we plot the QFI $\mathcal{H}$ and note that at the same values of $k$, the QFI is much larger, which ascertains that they can perform better sensing as well. We further explore how the signal $-D_Y$ varies with $V_Z$ to understand its response in Fig.~\ref{fig:Figure2}.\\
\indent The three values of the wave-vector $k$ where the the signal has a much higher proportionality with the Zeeman splitting is depicted in Fig.~\ref{fig:Figure2}(a). As we would expect for a small $V_Z$, the signal $D_Y$ shows a linear response which is captured in Fig.~\ref{fig:Figure2}(b) for the $k_1,k_2$ and $k_3$. However, for values of $V_Z>10^{-3}$ eV, it can be seen that the response stops being linear as can be noted from Fig.~\ref{fig:Figure2}(a). The value of $-D_Y$ actually begins to dip for $k_1$ after it hits the maximum possible value of 1. To understand the response in this range would require taking into account as the effects of higher orders of $\theta = V_Z/t_0$ in our signal \cite{2017Degen,dressel2012weak,Dressel2014}.\\
\indent We also compare the QFI with both the CFI and the maximum possible QFI in Fig.~\ref{fig:Figure3}. From these results, we can infer that at the resonant tunneling wave-vectors, $\mathcal{H}_c$ is closest to $\mathcal{H}$ which in turn is closest to the maximum value it can possibly attain (see  Fig.~\ref{fig:Figure2}(a) and Fig.~\ref{fig:Figure2}(c)). Another inference is that our modified barrier setup outperforms the single barrier setup by a very large margin at the resonant tunneling wave-vectors (see Fig.~\ref{fig:Figure2}(b) and Fig.~\ref{fig:Figure2}(d)). This shows that our sensor has the potential to give estimates with a near optimal error margin.
\subsection{Channels with dephasing}\label{subsec:dephas}
Solid-state systems are prone to dephasing interactions, typically categorized as pure-phase, phase and momentum and spin relaxations.  
The dephasing interactions that arise for pure-phase relaxation are usually electron-electron interactions. The interactions for momentum and phase relaxation are via fluctuating local non-magnetic impurities and that for spin relaxation are via magnetic impurities. These can be accounted for in the Keldysh NEGF method by adding the appropriate self-energies \cite{DANIELEWICZ1984239,datta1997electronic,book91179691,PhysRevB.75.081301,7571106,doi:10.1063/1.5023159,doi:10.1063/1.5044254,PhysRevApplied.8.064014,camsari2020,Duse_2021,PhysRevB.105.144418}.\\
\indent We define a scattering self-energy and the related in-scattering self-energy in the following matrix form \cite{camsari2020,PhysRevB.98.125417}
\begin{align}
    [\boldsymbol{\Sigma}^{r/<}_s]_{ij} &= D_{ijkl}[\mathbf{G}^{r/<}]_{kl}.
\end{align}
\begin{figure}[h]
    \centering
    \includegraphics[width=0.95\linewidth]{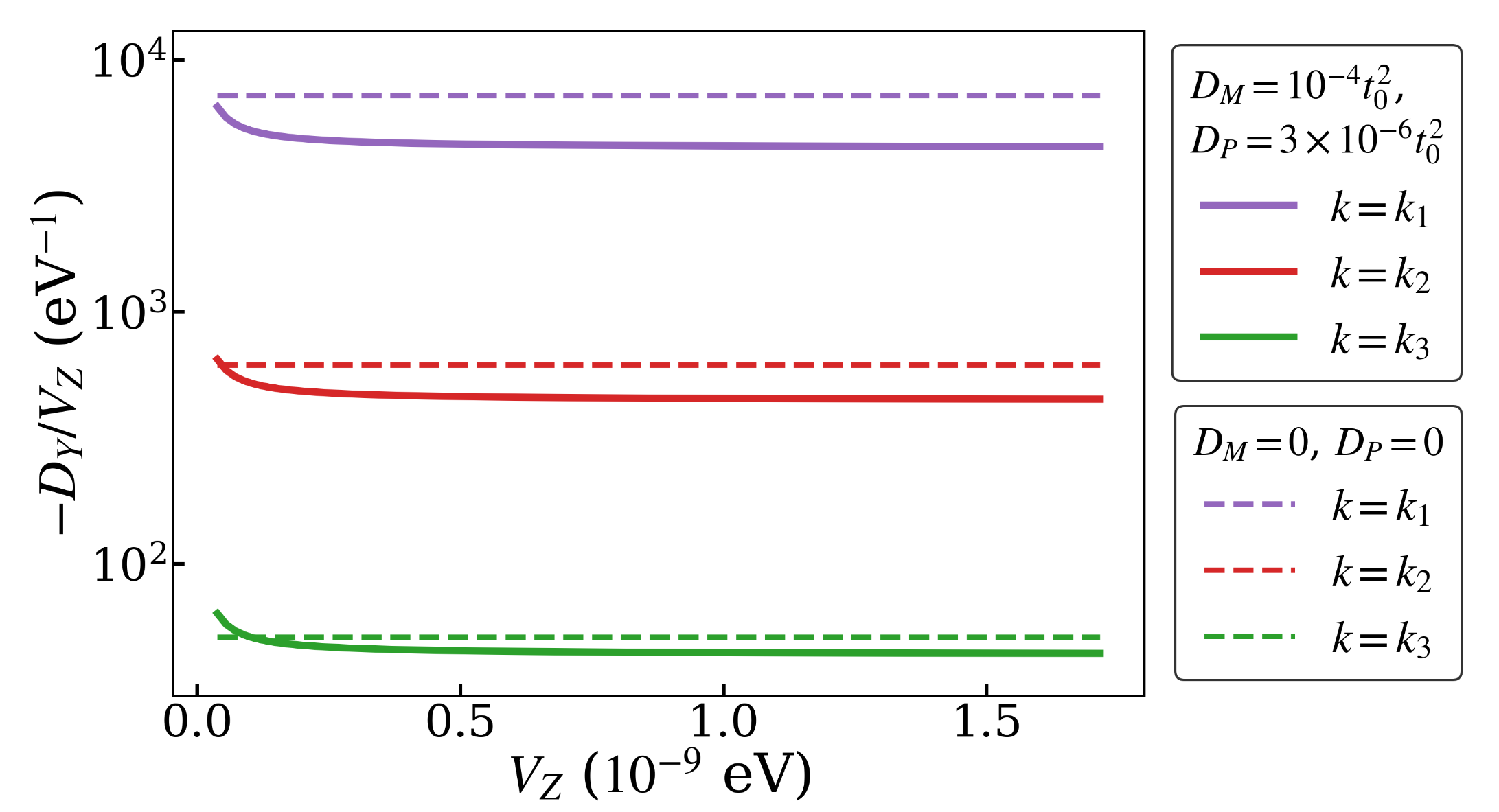}
    \caption{Magnetoresistance signal with pure-phase relaxation and momentum relaxation. The above graph shows the ratio between $D_Y$ and $V_Z$ showing that the response is not perfectly linear like in the absence of dephasing. They become almost perfectly linear following a certain value of $V_Z$.} \label{fig:Figure6}
\end{figure}
Here $D_{ijkl}$ is a rank-4 tensor which describes the spatial correlation between impurity scattering potentials \cite{camsari2020}. 
For pure-phase dephasing interactions, the tensor takes the following form characterized by interaction strength $D_P$ as
\begin{equation}
    D_{ijkl} = D_P\delta_{ik}\delta_{jl}.
\end{equation}
Here $\delta_{ij}$ is the Kronecker delta function. The corresponding tensor for momentum dephasing with strength $D_M$ is as follows
\begin{equation}
    D_{ijkl} = D_M\delta_{ij}\delta_{ik}\delta_{jl}.
\end{equation}
The self energies are then evaluated under the self-consistent Born approximation \cite{camsari2020}. It must be noted that both of these interactions do not affect spin, and hence do not affect the measurement setup. Accounting for spin dephasing effects will destroy the signal since the the setup is heavily dependent on spin coherence \cite{PhysRevB.105.144418}. Figure ~\ref{fig:Figure4} depicts the simulation results for both pure phase and momentum dephasing. Both of these effects broaden the peaks, as would be expected, but with important qualitative differences. Figure ~\ref{fig:Figure6} shows the results of simulating a channel with $D_M = 10^{-4}t_0^2$ and $D_P = 3\times10^{-6}t_0^2$, which correspond to typical impurity strengths encountered in 1-D channels. The linear behavior fails if we go below a Zeeman splitting less than $10^{-9}$ eV. 
There is a reduction in the slopes of this linear behavior compared to the slopes for the signals given by a clean channel. This reduction is not too large and still has the slope of the same order as can be deduced from Fig.~\ref{fig:Figure6}.

\section{Conclusion}
We proposed a spintronic device platform to realize weak-value enhanced quantum sensing. The setup estimates a very weak localized Zeeman splitting by exploiting a 
resonant tunneling enhanced magnetoresistance readout. We established that this paradigm offers a nearly optimal performance with a quantum Fisher information enhancement of about $10^4$ times that of single high-transmissivity barriers. The obtained signal also offers a high sensitivity in the presence of dephasing effects typically encountered in the solid state. These results, we believe, put forth definitive possibilities in harnessing the inherent sensitivity of resonant tunneling for solid-state quantum metrology with potential applications, especially, in the sensitive detection of small induced Zeeman effects \cite{dankert2017electrical,khokhriakov2020gate,kamalakar2016inversion}  in quantum material heterostructures. 

\section*{Acknowledgements} The authors acknowledge Kerem Camsari, Saroj Dash and Sai Vinjanampathy for useful discussions. The author BM wishes to acknowledge the financial support from the Science and Engineering Research Board (SERB), Government of India, under the MATRICS grant. 


\appendix
\section{1-D scattering and weak-values}\label{app:scatter}
We define the Hamiltonian of electrons in terms of spatial Hamiltonians $\hat{H}_0$ and $\hat{H}_1$ and some small dimensional parameter $\theta$ as
\begin{equation}
    \hat{H} = \hat{H}_0\otimes\mathbb{I} + \theta\hat{H}_1\otimes\sigma_Z.
\end{equation}
Let us look at the spectrum of scattering states with this Hamiltonian. We define a purely spatial scattering state $\ket{\psi}$ as follows
\begin{equation}
    \hat{H}_0\ket{\psi} = \varepsilon\ket{\psi}.
\end{equation}
This will get scattered further due to the $\theta H_1\otimes\sigma_Z$ part. Let us now define $\ket{\psi_\pm}$ as the spatially scattered states for the up-spin and the down-spin channels respectively, expressed as
\begin{equation}
    \ket{\psi_\pm} = \ket{\psi} \pm \dfrac{\theta\hat{H}_1}{\varepsilon - \hat{H}_0}\ket{\psi_\pm} = \ket{\psi} \pm \hat{G}_0(\varepsilon)\theta\hat{H}_1\ket{\psi_\pm}.
\end{equation}
Here $\hat{G}_0$ is the equilibrium isolated Green's function of the Hamiltonian $\hat{H}_0$. By the general convention, the Green's function is defined as $\hat{G}_0(\varepsilon) = [\varepsilon - H_0 \pm i\eta]^{-1}$ with the plus and minus choice representing the retarded or the advanced Green's function. This choice will be largely irrelevant to how we use this operator since it is never acted on an eigenstate directly. For the sake of convention, all mentions of $\hat{G}_0$ will be of the retarded Green's function which has a more physical relevance \cite{sakurai_napolitano_2017}.\\
\indent We now define $\ket{f}$ as a part of $\ket{\psi}$ which is after scattering  \cite{Steinberg1995}. If we assume that the incident wave is $\ket{\psi}\otimes\ket{+\hat{x}}$, the scattered wave is $(\ket{\psi_+}\ket{+\hat{z}} + \ket{\psi_-}\ket{-\hat{z}})/\sqrt{2}$. We can evaluate the expectation value $\langle\sigma_y\rangle$ for the part on the left of the scattering section (including barriers in $\hat{H}_0$) as follows
\begin{equation}\label{eq:Sy}
    \langle\sigma_Y\rangle = \dfrac{i\langle\psi_-|f\rangle\langle f|\psi_+\rangle - i\langle\psi_+|f\rangle\langle f|\psi_-\rangle}{|\langle\psi_+|f\rangle|^2 + |\langle\psi_-|f\rangle|^2}.
\end{equation}
The problem of 1-D scattering has been dealt with in more depth in Ref. \cite{Aharonov2008}. To make a qualitative argument of the proportionality to the weak-value, we can choose to use the Born approximation. This gives $\ket{\psi_\pm} \approx \ket{\psi} \pm \theta\hat{G}_0\hat{H}_1\ket{\psi}$ which then helps to simplify \eqref{eq:Sy} to the following form
\begin{equation}\begin{aligned}
    \langle\sigma_Y\rangle &= i\theta\frac{\langle \psi|f\rangle\langle f|\hat{G}_0 \hat{H}_1|\psi\rangle - \langle \psi|\hat{H}_1\hat{G}^\dagger_0|f\rangle\langle f|\psi\rangle}{\langle\psi|f\rangle\langle f|\psi\rangle + \theta^2\langle\psi|\hat{H}_1\hat{G}^\dagger_0|f\rangle\langle f|\hat{G}_0 \hat{H}_1|\psi\rangle} \\
    &= -2\theta\Im\left(\dfrac{\langle f|\hat{G}_0 \hat{H}_1|\psi\rangle}{\langle f|\psi\rangle}\right) + \mathcal{O}(\theta^2).
\end{aligned}\end{equation}
Notably the first order term is simply the weak-value of $\hat{G}_0\hat{H}_1$. To actually evaluate this, we must note that $\hat{H}_1$ is only non-zero in the region of the middle barrier. We can then act the Green's function on $\bra{f}$ and we will finally get an integral which is only in the spatial region of the middle barrier as described in \cite{Steinberg1995}.\\
\indent An important insight this calculation gives us is that due to the form of $\hat{G}_0\hat{H}_1$ for 1-D barriers, we can have a case where the weak-value has an amplified value for the choice of $\ket{\psi}$ which has full transmission (hence maximum $\langle f|\psi\rangle$)as is observed in our results as well.
\section{Quantum Fisher information for the setup}\label{app:QFI}
In this section, we will work out the expression for the quantum fisher information which we can obtain out of or resonant tunneling setup. The Hamiltonian is defined in equation \eqref{eq:Hamiltonian}. We wish to estimate $\theta$ to measure Zeeman splitting. This is also a problem that has been studied in context of quantum walks for 1-D scattering \cite{e22111321}. We define the following position Hamiltonians
\begin{equation}\hat{H}_0 = \begin{cases}\frac{p^2}{2m} + V_{B1} & |y|\leq \frac{d_1}{2}\\\frac{p^2}{2m} + V_{B2} & \frac{d_1}{2}<|y|\leq \frac{d_2}{2}\\\frac{p^2}{2m} & |y|>\frac{d_2}{2}\end{cases},\end{equation}
\begin{equation}\hat{H}_1 = \begin{cases}t_0 & |y| \leq \frac{d_1}{2}\\0 & |y| > \frac{d_1}{2}\end{cases}.\end{equation}
For spin up (or spin down) particles, the effective Hamiltonian is $H_0 + \theta H_1$ (or $H_0 - \theta H_1$). One of the defined Green's function based on the number of particles in the channel is the $G^n$ function defined in terms of the advanced and retarded Green's functions as $\mathbf{G}^n = \mathbf{G}^r\boldsymbol{\Sigma}_{in}\mathbf{G}^a$. We obtain, $\rho = \mathbf{G}^n/\text{Tr}(\mathbf{G}^n)$ and so we can see the following on taking a partial derivative with respect to our parameter:
\begin{equation}\frac{\partial\mathbf{G}^n}{\partial \theta} = \frac{\partial\mathbf{G}^r}{\partial \theta}\boldsymbol{\Sigma}_{in}\mathbf{G}^a + \mathbf{G}^r\boldsymbol{\Sigma}_{in}\frac{\partial \mathbf{G}a}{\partial \theta}\end{equation}
Now we must note that the retarded green's function is defined as follows
\begin{equation}\label{eq:Gr}\mathbf{G}^r = [(E+i\eta)\mathbb{I} - H_0\otimes\mathbb{I}_2 - \theta H_1\otimes\sigma_z - \boldsymbol{\Sigma}_L - \boldsymbol{\Sigma}_R - \boldsymbol{\Sigma}_{F1} - \boldsymbol{\Sigma}_{F2}]^{-1}.\end{equation}
From this we can find the partial derivative of $\mathbf{G}^r$ with respect to the parameter $\theta$.
\begin{equation}\frac{\partial\mathbf{G}^r}{\partial \theta} = -H_1\times-(\mathbf{G}^r)^2 = H_1\mathbf{G}^r\times \mathbf{G}^r\end{equation}
Hence if we define $L = H_1G^r$ we can clearly see that the following holds
\begin{equation}\label{eq:derGr}\frac{\partial\mathbf{G}^n}{\partial\theta} = L\mathbf{G}^n + \mathbf{G}^nL^\dagger\end{equation}
We must now also account for the fact that $G^n$ must be normalized to give the expression of the density matrix.
\begin{align}
\begin{split}
    \frac{\partial\rho}{\partial\theta} &= \frac{\partial}{\partial\theta}\frac{\mathbf{G}^n}{\text{Tr}(\mathbf{G}^n)} \\
    &= \frac{1}{\text{Tr}(\mathbf{G}^n)}\frac{\partial\mathbf{G}^n}{\partial\theta} - \frac{\mathbf{G}^n}{\text{Tr}(\mathbf{G}^n)^2}\text{Tr}\left(\frac{\partial \mathbf{G}^n}{\partial\theta}\right)\\
    &= \left(L - \frac{\text{Tr}(L\mathbf{G}^n)}{\text{Tr}(\mathbf{G}^n)}\mathbb{I}\right)\rho + \rho\left(L^\dagger - \frac{\text{Tr}(\mathbf{G}^nL^\dagger)}{\text{Tr}(\mathbf{G}^n)}\mathbb{I}\right)
\end{split}
\end{align}
$Tr(A^\dagger)\mathbb{I} = ((Tr(A))\mathbb{I})^\dagger$ hence if we define $\tilde{L}$ as follows:
\begin{equation}\tilde{L} = L - \frac{\text{Tr}(L\mathbf{G}^n)}{\text{Tr}(\mathbf{G}^n)}\mathbb{I},\end{equation}
we can write the following expression
\begin{equation}\partial_\theta\rho = \tilde{L}\rho + \rho\tilde{L}^\dagger.\end{equation}
This may look a lot like the expression of QFI defined in terms of a symmetric logarithmic derivative \cite{Liu_2019}. The operator $\tilde{L}$ isn't Hermitian hence fails to be a symmetric logarithmic derivative. In general QFI is defined as the following for density matrix $\rho = \sum \lambda_i\ket{i}\bra{i}$
\begin{equation}\mathcal{H} = \sum_{i,j, \lambda_i+\lambda_j\neq0}2\text{Re}\frac{\braket{i|\partial_\theta\rho|i}\braket{j|\partial_\theta\rho|j}}{\lambda_i+\lambda_j}\end{equation}
If $\partial_\theta\rho = \tilde{L}\rho + \rho\tilde{L}^\dagger$, then $\braket{i|\partial_\theta\rho|i} = \lambda_i\braket{i|\tilde{L}+\tilde{L}^\dagger|i}$. Hence if $\rho$ is pure we get the following (let  $\rho = \ket{\psi}\bra{\psi}$).
\begin{equation}
\begin{aligned}
    \mathcal{H} &= \text{Re}\left(\braket{\psi|\tilde{L} + \tilde{L}^\dagger|\psi}^2\right)\\ &= \text{Tr}(\rho(\tilde{L}+\tilde{L}^\dagger)^2)
\end{aligned}\end{equation}
It is a known result that QFI maximizes when we have $\rho_\theta$ being a pure state.\\
It is easy to see that based on this we have
\begin{equation}\mathcal{H} \leq \max(\text{eigenvalues}((\tilde{L}+\tilde{L}^\dagger)^2)).\end{equation}
\section{Current measurement as a strong measurement}\label{app:measure}
The act of obtaining currents at the ferromagnetic contacts gives the statistics for the spin expectation values. This is due to the fact that the ferromagnetic contact, if aligned along a certain direction, will give a current readout proportional to the population of spins aligned in that particular direction \cite{PhysRevB.105.144418}. This can be established in the NEGF formulation. The current readouts are from the $FM2$ contact at $\pm\hat{y}$ orientation. The current values come out to be as follows
\begin{equation}
\begin{aligned}
    I^\pm_{FM2} &= \text{Tr}(\Gamma^{FM2}\mathbf{G}^n)\\ &= -2t_0i\sin(ka) \text{Tr}(\mathbf{G}^n)\text{Tr}\left(\frac{(\mathbb{I}\pm\sigma_y)\delta_{f_2,f_2}}{2}\rho\right).
\end{aligned}
\end{equation}
The quantity $\text{Tr}((\mathbb{I}\pm\sigma_y)\rho/2)$ is simply the probabilities for the POVM set of $\{(\mathbb{I}+\sigma_y)/2,(\mathbb{I}-\sigma_y)/2\}$. An additional point to note is that our post-selection measurement is looking at one point in the whole region which lies to the right of the barrier region (electrons are injected from the left of the barrier). The reason it is only one point is since the current readout only occurs at a specific point in the 1-D nanowire. This has no change on the expectation value of $\sigma_Y$ since this will have to be same all over the whole region which lies to the right of the barrier region.\\
\indent Hence what we use as the expectation value of $\sigma_Y$ is the same as the expression we obtain by considering a complete post-selected region in equation \eqref{eq:Sy} since the spin part of the wavefunction is the same everywhere on the right of the barrier. For any 1-D scattering problem, all changes only occur at the boundaries, hence by looking at one point we can get the relevant information for the whole post-selected region.
\section{Classical Fisher information for the setup}\label{app:CFI}
As we have established previously, we take the current readouts to behave as probabilities for the POVM set of $\{(\mathbb{I}+\sigma_y)/2,(\mathbb{I}-\sigma_y)/2\}$
There are a few issues with taking this as a direct interpretation since the state $\rho$ is ultimately dependent on the polarization of $FM2$ (see equation \eqref{eq:Gr}) hence is slightly different depending on whether it is $+\hat{y}$ or $-\hat{y}$. We first define the probabilities $p_\pm$ as
\begin{align}
    p_\pm = \frac{\text{Tr}\left(\frac{(\mathbb{I}\pm\sigma_y)\delta_{f_2,f_2}}{2}\mathbf{G}^n_\pm\right)}{\text{Tr}(\mathbf{G}^n_\pm)}.
\end{align}
We must note that these probabilities are only looking at a certain lattice point corresponding to the contact $FM2$. Hence we actually need to define conditional properties since those are the actual probabilities we get $\pm\hat{y}$ polarization electrons detected on the other end. Hence let $\tilde{p}_\pm = p_\pm/(p_+ + p_-)$. From this, the CFI is simply given as follows.
\begin{equation}
    \mathcal{H}_c = \dfrac{(\partial_\theta \tilde{p}_+)^2}{\tilde{p}_+} + \dfrac{(\partial_\theta \tilde{p}_-)^2}{\tilde{p}_-}
\end{equation}
The expression of this can be easily evaluated using equation \eqref{eq:derGr}.

\bibliography{references.bib}

\begin{thebibliography}{80}%
\makeatletter
\providecommand \@ifxundefined [1]{%
 \@ifx{#1\undefined}
}%
\providecommand \@ifnum [1]{%
 \ifnum #1\expandafter \@firstoftwo
 \else \expandafter \@secondoftwo
 \fi
}%
\providecommand \@ifx [1]{%
 \ifx #1\expandafter \@firstoftwo
 \else \expandafter \@secondoftwo
 \fi
}%
\providecommand \natexlab [1]{#1}%
\providecommand \enquote  [1]{``#1''}%
\providecommand \bibnamefont  [1]{#1}%
\providecommand \bibfnamefont [1]{#1}%
\providecommand \citenamefont [1]{#1}%
\providecommand \href@noop [0]{\@secondoftwo}%
\providecommand \href [0]{\begingroup \@sanitize@url \@href}%
\providecommand \@href[1]{\@@startlink{#1}\@@href}%
\providecommand \@@href[1]{\endgroup#1\@@endlink}%
\providecommand \@sanitize@url [0]{\catcode `\\12\catcode `\$12\catcode
  `\&12\catcode `\#12\catcode `\^12\catcode `\_12\catcode `\%12\relax}%
\providecommand \@@startlink[1]{}%
\providecommand \@@endlink[0]{}%
\providecommand \url  [0]{\begingroup\@sanitize@url \@url }%
\providecommand \@url [1]{\endgroup\@href {#1}{\urlprefix }}%
\providecommand \urlprefix  [0]{URL }%
\providecommand \Eprint [0]{\href }%
\providecommand \doibase [0]{https://doi.org/}%
\providecommand \selectlanguage [0]{\@gobble}%
\providecommand \bibinfo  [0]{\@secondoftwo}%
\providecommand \bibfield  [0]{\@secondoftwo}%
\providecommand \translation [1]{[#1]}%
\providecommand \BibitemOpen [0]{}%
\providecommand \bibitemStop [0]{}%
\providecommand \bibitemNoStop [0]{.\EOS\space}%
\providecommand \EOS [0]{\spacefactor3000\relax}%
\providecommand \BibitemShut  [1]{\csname bibitem#1\endcsname}%
\let\auto@bib@innerbib\@empty
\bibitem [{\citenamefont {Giovannetti}\ \emph {et~al.}(2006)\citenamefont
  {Giovannetti}, \citenamefont {Lloyd},\ and\ \citenamefont
  {Maccone}}]{PhysRevLett.96.010401}%
  \BibitemOpen
  \bibfield  {author} {\bibinfo {author} {\bibfnamefont {V.}~\bibnamefont
  {Giovannetti}}, \bibinfo {author} {\bibfnamefont {S.}~\bibnamefont {Lloyd}},\
  and\ \bibinfo {author} {\bibfnamefont {L.}~\bibnamefont {Maccone}},\ }\href
  {https://doi.org/10.1103/PhysRevLett.96.010401} {\bibfield  {journal}
  {\bibinfo  {journal} {Phys. Rev. Lett.}\ }\textbf {\bibinfo {volume} {96}},\
  \bibinfo {pages} {010401} (\bibinfo {year} {2006})}\BibitemShut {NoStop}%
\bibitem [{\citenamefont {Giovannetti}\ \emph {et~al.}(2011)\citenamefont
  {Giovannetti}, \citenamefont {Lloyd},\ and\ \citenamefont
  {Maccone}}]{giovannetti2011advances}%
  \BibitemOpen
  \bibfield  {author} {\bibinfo {author} {\bibfnamefont {V.}~\bibnamefont
  {Giovannetti}}, \bibinfo {author} {\bibfnamefont {S.}~\bibnamefont {Lloyd}},\
  and\ \bibinfo {author} {\bibfnamefont {L.}~\bibnamefont {Maccone}},\
  }\href@noop {} {\bibfield  {journal} {\bibinfo  {journal} {Nature photonics}\
  }\textbf {\bibinfo {volume} {5}},\ \bibinfo {pages} {222} (\bibinfo {year}
  {2011})}\BibitemShut {NoStop}%
\bibitem [{\citenamefont {Degen}\ \emph {et~al.}(2017)\citenamefont {Degen},
  \citenamefont {Reinhard},\ and\ \citenamefont {Cappellaro}}]{2017Degen}%
  \BibitemOpen
  \bibfield  {author} {\bibinfo {author} {\bibfnamefont {C.}~\bibnamefont
  {Degen}}, \bibinfo {author} {\bibfnamefont {F.}~\bibnamefont {Reinhard}},\
  and\ \bibinfo {author} {\bibfnamefont {P.}~\bibnamefont {Cappellaro}},\
  }\bibfield  {journal} {\bibinfo  {journal} {Reviews of Modern Physics}\
  }\textbf {\bibinfo {volume} {89}},\ \href
  {https://doi.org/10.1103/revmodphys.89.035002} {10.1103/revmodphys.89.035002}
  (\bibinfo {year} {2017})\BibitemShut {NoStop}%
\bibitem [{\citenamefont {Polino}\ \emph {et~al.}(2020)\citenamefont {Polino},
  \citenamefont {Valeri}, \citenamefont {Spagnolo},\ and\ \citenamefont
  {Sciarrino}}]{doi:10.1116/5.0007577}%
  \BibitemOpen
  \bibfield  {author} {\bibinfo {author} {\bibfnamefont {E.}~\bibnamefont
  {Polino}}, \bibinfo {author} {\bibfnamefont {M.}~\bibnamefont {Valeri}},
  \bibinfo {author} {\bibfnamefont {N.}~\bibnamefont {Spagnolo}},\ and\
  \bibinfo {author} {\bibfnamefont {F.}~\bibnamefont {Sciarrino}},\ }\href
  {https://doi.org/10.1116/5.0007577} {\bibfield  {journal} {\bibinfo
  {journal} {AVS Quantum Science}\ }\textbf {\bibinfo {volume} {2}},\ \bibinfo
  {pages} {024703} (\bibinfo {year} {2020})},\ \Eprint
  {https://arxiv.org/abs/https://doi.org/10.1116/5.0007577}
  {https://doi.org/10.1116/5.0007577} \BibitemShut {NoStop}%
\bibitem [{\citenamefont {Taylor}\ and\ \citenamefont
  {Bowen}(2016)}]{TAYLOR20161}%
  \BibitemOpen
  \bibfield  {author} {\bibinfo {author} {\bibfnamefont {M.~A.}\ \bibnamefont
  {Taylor}}\ and\ \bibinfo {author} {\bibfnamefont {W.~P.}\ \bibnamefont
  {Bowen}},\ }\href
  {https://doi.org/https://doi.org/10.1016/j.physrep.2015.12.002} {\bibfield
  {journal} {\bibinfo  {journal} {Physics Reports}\ }\textbf {\bibinfo {volume}
  {615}},\ \bibinfo {pages} {1} (\bibinfo {year} {2016})},\ \bibinfo {note}
  {quantum metrology and its application in biology}\BibitemShut {NoStop}%
\bibitem [{\citenamefont {Joo}\ \emph {et~al.}(2011)\citenamefont {Joo},
  \citenamefont {Munro},\ and\ \citenamefont
  {Spiller}}]{PhysRevLett.107.083601}%
  \BibitemOpen
  \bibfield  {author} {\bibinfo {author} {\bibfnamefont {J.}~\bibnamefont
  {Joo}}, \bibinfo {author} {\bibfnamefont {W.~J.}\ \bibnamefont {Munro}},\
  and\ \bibinfo {author} {\bibfnamefont {T.~P.}\ \bibnamefont {Spiller}},\
  }\href {https://doi.org/10.1103/PhysRevLett.107.083601} {\bibfield  {journal}
  {\bibinfo  {journal} {Phys. Rev. Lett.}\ }\textbf {\bibinfo {volume} {107}},\
  \bibinfo {pages} {083601} (\bibinfo {year} {2011})}\BibitemShut {NoStop}%
\bibitem [{\citenamefont {Pang}\ and\ \citenamefont
  {Brun}(2014)}]{PhysRevA.90.022117}%
  \BibitemOpen
  \bibfield  {author} {\bibinfo {author} {\bibfnamefont {S.}~\bibnamefont
  {Pang}}\ and\ \bibinfo {author} {\bibfnamefont {T.~A.}\ \bibnamefont
  {Brun}},\ }\href {https://doi.org/10.1103/PhysRevA.90.022117} {\bibfield
  {journal} {\bibinfo  {journal} {Phys. Rev. A}\ }\textbf {\bibinfo {volume}
  {90}},\ \bibinfo {pages} {022117} (\bibinfo {year} {2014})}\BibitemShut
  {NoStop}%
\bibitem [{\citenamefont {Kaubruegger}\ \emph {et~al.}(2021)\citenamefont
  {Kaubruegger}, \citenamefont {Vasilyev}, \citenamefont {Schulte},
  \citenamefont {Hammerer},\ and\ \citenamefont {Zoller}}]{PhysRevX.11.041045}%
  \BibitemOpen
  \bibfield  {author} {\bibinfo {author} {\bibfnamefont {R.}~\bibnamefont
  {Kaubruegger}}, \bibinfo {author} {\bibfnamefont {D.~V.}\ \bibnamefont
  {Vasilyev}}, \bibinfo {author} {\bibfnamefont {M.}~\bibnamefont {Schulte}},
  \bibinfo {author} {\bibfnamefont {K.}~\bibnamefont {Hammerer}},\ and\
  \bibinfo {author} {\bibfnamefont {P.}~\bibnamefont {Zoller}},\ }\href
  {https://doi.org/10.1103/PhysRevX.11.041045} {\bibfield  {journal} {\bibinfo
  {journal} {Phys. Rev. X}\ }\textbf {\bibinfo {volume} {11}},\ \bibinfo
  {pages} {041045} (\bibinfo {year} {2021})}\BibitemShut {NoStop}%
\bibitem [{\citenamefont {Marciniak}\ \emph {et~al.}(2022)\citenamefont
  {Marciniak}, \citenamefont {Feldker}, \citenamefont {Pogorelov},
  \citenamefont {Kaubruegger}, \citenamefont {Vasilyev}, \citenamefont {van
  Bijnen}, \citenamefont {Schindler}, \citenamefont {Zoller}, \citenamefont
  {Blatt},\ and\ \citenamefont {Monz}}]{Marciniak_2022}%
  \BibitemOpen
  \bibfield  {author} {\bibinfo {author} {\bibfnamefont {C.~D.}\ \bibnamefont
  {Marciniak}}, \bibinfo {author} {\bibfnamefont {T.}~\bibnamefont {Feldker}},
  \bibinfo {author} {\bibfnamefont {I.}~\bibnamefont {Pogorelov}}, \bibinfo
  {author} {\bibfnamefont {R.}~\bibnamefont {Kaubruegger}}, \bibinfo {author}
  {\bibfnamefont {D.~V.}\ \bibnamefont {Vasilyev}}, \bibinfo {author}
  {\bibfnamefont {R.}~\bibnamefont {van Bijnen}}, \bibinfo {author}
  {\bibfnamefont {P.}~\bibnamefont {Schindler}}, \bibinfo {author}
  {\bibfnamefont {P.}~\bibnamefont {Zoller}}, \bibinfo {author} {\bibfnamefont
  {R.}~\bibnamefont {Blatt}},\ and\ \bibinfo {author} {\bibfnamefont
  {T.}~\bibnamefont {Monz}},\ }\href
  {https://doi.org/10.1038/s41586-022-04435-4} {\bibfield  {journal} {\bibinfo
  {journal} {Nature}\ }\textbf {\bibinfo {volume} {603}},\ \bibinfo {pages}
  {604} (\bibinfo {year} {2022})}\BibitemShut {NoStop}%
\bibitem [{\citenamefont {Hofmann}(2011)}]{2011Hofmann}%
  \BibitemOpen
  \bibfield  {author} {\bibinfo {author} {\bibfnamefont {H.~F.}\ \bibnamefont
  {Hofmann}},\ }\bibfield  {journal} {\bibinfo  {journal} {Physical Review A}\
  }\textbf {\bibinfo {volume} {83}},\ \href
  {https://doi.org/10.1103/physreva.83.022106} {10.1103/physreva.83.022106}
  (\bibinfo {year} {2011})\BibitemShut {NoStop}%
\bibitem [{\citenamefont {Kofman}\ \emph {et~al.}(2012)\citenamefont {Kofman},
  \citenamefont {Ashhab},\ and\ \citenamefont {Nori}}]{2012Kofman}%
  \BibitemOpen
  \bibfield  {author} {\bibinfo {author} {\bibfnamefont {A.~G.}\ \bibnamefont
  {Kofman}}, \bibinfo {author} {\bibfnamefont {S.}~\bibnamefont {Ashhab}},\
  and\ \bibinfo {author} {\bibfnamefont {F.}~\bibnamefont {Nori}},\ }\href
  {https://doi.org/10.1016/j.physrep.2012.07.001} {\bibfield  {journal}
  {\bibinfo  {journal} {Physics Reports}\ }\textbf {\bibinfo {volume} {520}},\
  \bibinfo {pages} {43–133} (\bibinfo {year} {2012})}\BibitemShut {NoStop}%
\bibitem [{\citenamefont {Dressel}\ and\ \citenamefont
  {Jordan}(2012)}]{dressel2012weak}%
  \BibitemOpen
  \bibfield  {author} {\bibinfo {author} {\bibfnamefont {J.}~\bibnamefont
  {Dressel}}\ and\ \bibinfo {author} {\bibfnamefont {A.~N.}\ \bibnamefont
  {Jordan}},\ }\href@noop {} {\bibfield  {journal} {\bibinfo  {journal}
  {Physical review letters}\ }\textbf {\bibinfo {volume} {109}},\ \bibinfo
  {pages} {230402} (\bibinfo {year} {2012})}\BibitemShut {NoStop}%
\bibitem [{\citenamefont {Lyons}\ \emph {et~al.}(2015)\citenamefont {Lyons},
  \citenamefont {Dressel}, \citenamefont {Jordan}, \citenamefont {Howell},\
  and\ \citenamefont {Kwiat}}]{PhysRevLett.114.170801}%
  \BibitemOpen
  \bibfield  {author} {\bibinfo {author} {\bibfnamefont {K.}~\bibnamefont
  {Lyons}}, \bibinfo {author} {\bibfnamefont {J.}~\bibnamefont {Dressel}},
  \bibinfo {author} {\bibfnamefont {A.~N.}\ \bibnamefont {Jordan}}, \bibinfo
  {author} {\bibfnamefont {J.~C.}\ \bibnamefont {Howell}},\ and\ \bibinfo
  {author} {\bibfnamefont {P.~G.}\ \bibnamefont {Kwiat}},\ }\href
  {https://doi.org/10.1103/PhysRevLett.114.170801} {\bibfield  {journal}
  {\bibinfo  {journal} {Phys. Rev. Lett.}\ }\textbf {\bibinfo {volume} {114}},\
  \bibinfo {pages} {170801} (\bibinfo {year} {2015})}\BibitemShut {NoStop}%
\bibitem [{\citenamefont {Viza}\ \emph {et~al.}(2015)\citenamefont {Viza},
  \citenamefont {Mart\'{\i}nez-Rinc\'on}, \citenamefont {Alves}, \citenamefont
  {Jordan},\ and\ \citenamefont {Howell}}]{PhysRevA.92.032127}%
  \BibitemOpen
  \bibfield  {author} {\bibinfo {author} {\bibfnamefont {G.~I.}\ \bibnamefont
  {Viza}}, \bibinfo {author} {\bibfnamefont {J.}~\bibnamefont
  {Mart\'{\i}nez-Rinc\'on}}, \bibinfo {author} {\bibfnamefont {G.~B.}\
  \bibnamefont {Alves}}, \bibinfo {author} {\bibfnamefont {A.~N.}\ \bibnamefont
  {Jordan}},\ and\ \bibinfo {author} {\bibfnamefont {J.~C.}\ \bibnamefont
  {Howell}},\ }\href {https://doi.org/10.1103/PhysRevA.92.032127} {\bibfield
  {journal} {\bibinfo  {journal} {Phys. Rev. A}\ }\textbf {\bibinfo {volume}
  {92}},\ \bibinfo {pages} {032127} (\bibinfo {year} {2015})}\BibitemShut
  {NoStop}%
\bibitem [{\citenamefont {Xu}\ \emph {et~al.}(2020)\citenamefont {Xu},
  \citenamefont {Liu}, \citenamefont {Datta}, \citenamefont {Knee},
  \citenamefont {Lundeen}, \citenamefont {Lu},\ and\ \citenamefont
  {Zhang}}]{PhysRevLett.125.080501}%
  \BibitemOpen
  \bibfield  {author} {\bibinfo {author} {\bibfnamefont {L.}~\bibnamefont
  {Xu}}, \bibinfo {author} {\bibfnamefont {Z.}~\bibnamefont {Liu}}, \bibinfo
  {author} {\bibfnamefont {A.}~\bibnamefont {Datta}}, \bibinfo {author}
  {\bibfnamefont {G.~C.}\ \bibnamefont {Knee}}, \bibinfo {author}
  {\bibfnamefont {J.~S.}\ \bibnamefont {Lundeen}}, \bibinfo {author}
  {\bibfnamefont {Y.-q.}\ \bibnamefont {Lu}},\ and\ \bibinfo {author}
  {\bibfnamefont {L.}~\bibnamefont {Zhang}},\ }\href
  {https://doi.org/10.1103/PhysRevLett.125.080501} {\bibfield  {journal}
  {\bibinfo  {journal} {Phys. Rev. Lett.}\ }\textbf {\bibinfo {volume} {125}},\
  \bibinfo {pages} {080501} (\bibinfo {year} {2020})}\BibitemShut {NoStop}%
\bibitem [{\citenamefont {Liu}\ \emph {et~al.}(2022)\citenamefont {Liu},
  \citenamefont {Qin},\ and\ \citenamefont {Li}}]{PhysRevA.106.022619}%
  \BibitemOpen
  \bibfield  {author} {\bibinfo {author} {\bibfnamefont {Y.}~\bibnamefont
  {Liu}}, \bibinfo {author} {\bibfnamefont {L.}~\bibnamefont {Qin}},\ and\
  \bibinfo {author} {\bibfnamefont {X.-Q.}\ \bibnamefont {Li}},\ }\href
  {https://doi.org/10.1103/PhysRevA.106.022619} {\bibfield  {journal} {\bibinfo
   {journal} {Phys. Rev. A}\ }\textbf {\bibinfo {volume} {106}},\ \bibinfo
  {pages} {022619} (\bibinfo {year} {2022})}\BibitemShut {NoStop}%
\bibitem [{\citenamefont {Liu}\ \emph {et~al.}(2019)\citenamefont {Liu},
  \citenamefont {Yuan}, \citenamefont {Lu},\ and\ \citenamefont
  {Wang}}]{Liu_2019}%
  \BibitemOpen
  \bibfield  {author} {\bibinfo {author} {\bibfnamefont {J.}~\bibnamefont
  {Liu}}, \bibinfo {author} {\bibfnamefont {H.}~\bibnamefont {Yuan}}, \bibinfo
  {author} {\bibfnamefont {X.-M.}\ \bibnamefont {Lu}},\ and\ \bibinfo {author}
  {\bibfnamefont {X.}~\bibnamefont {Wang}},\ }\href
  {https://doi.org/10.1088/1751-8121/ab5d4d} {\bibfield  {journal} {\bibinfo
  {journal} {Journal of Physics A: Mathematical and Theoretical}\ }\textbf
  {\bibinfo {volume} {53}},\ \bibinfo {pages} {023001} (\bibinfo {year}
  {2019})}\BibitemShut {NoStop}%
\bibitem [{\citenamefont {Paris}(2009)}]{paris2009quantum}%
  \BibitemOpen
  \bibfield  {author} {\bibinfo {author} {\bibfnamefont {M.~G.}\ \bibnamefont
  {Paris}},\ }\href@noop {} {\bibfield  {journal} {\bibinfo  {journal}
  {International Journal of Quantum Information}\ }\textbf {\bibinfo {volume}
  {7}},\ \bibinfo {pages} {125} (\bibinfo {year} {2009})}\BibitemShut {NoStop}%
\bibitem [{\citenamefont {Facchi}\ \emph {et~al.}(2010)\citenamefont {Facchi},
  \citenamefont {Kulkarni}, \citenamefont {Man'ko}, \citenamefont {Marmo},
  \citenamefont {Sudarshan},\ and\ \citenamefont
  {Ventriglia}}]{FACCHI20104801}%
  \BibitemOpen
  \bibfield  {author} {\bibinfo {author} {\bibfnamefont {P.}~\bibnamefont
  {Facchi}}, \bibinfo {author} {\bibfnamefont {R.}~\bibnamefont {Kulkarni}},
  \bibinfo {author} {\bibfnamefont {V.}~\bibnamefont {Man'ko}}, \bibinfo
  {author} {\bibfnamefont {G.}~\bibnamefont {Marmo}}, \bibinfo {author}
  {\bibfnamefont {E.}~\bibnamefont {Sudarshan}},\ and\ \bibinfo {author}
  {\bibfnamefont {F.}~\bibnamefont {Ventriglia}},\ }\href
  {https://doi.org/https://doi.org/10.1016/j.physleta.2010.10.005} {\bibfield
  {journal} {\bibinfo  {journal} {Physics Letters A}\ }\textbf {\bibinfo
  {volume} {374}},\ \bibinfo {pages} {4801} (\bibinfo {year}
  {2010})}\BibitemShut {NoStop}%
\bibitem [{\citenamefont {Fujiwara}\ and\ \citenamefont
  {Nagaoka}(1995)}]{FUJIWARA1995119}%
  \BibitemOpen
  \bibfield  {author} {\bibinfo {author} {\bibfnamefont {A.}~\bibnamefont
  {Fujiwara}}\ and\ \bibinfo {author} {\bibfnamefont {H.}~\bibnamefont
  {Nagaoka}},\ }\href
  {https://doi.org/https://doi.org/10.1016/0375-9601(95)00269-9} {\bibfield
  {journal} {\bibinfo  {journal} {Physics Letters A}\ }\textbf {\bibinfo
  {volume} {201}},\ \bibinfo {pages} {119} (\bibinfo {year}
  {1995})}\BibitemShut {NoStop}%
\bibitem [{\citenamefont {Alves}\ \emph {et~al.}(2015)\citenamefont {Alves},
  \citenamefont {Escher}, \citenamefont {de~Matos~Filho}, \citenamefont
  {Zagury},\ and\ \citenamefont {Davidovich}}]{PhysRevA.91.062107}%
  \BibitemOpen
  \bibfield  {author} {\bibinfo {author} {\bibfnamefont {G.~B.}\ \bibnamefont
  {Alves}}, \bibinfo {author} {\bibfnamefont {B.~M.}\ \bibnamefont {Escher}},
  \bibinfo {author} {\bibfnamefont {R.~L.}\ \bibnamefont {de~Matos~Filho}},
  \bibinfo {author} {\bibfnamefont {N.}~\bibnamefont {Zagury}},\ and\ \bibinfo
  {author} {\bibfnamefont {L.}~\bibnamefont {Davidovich}},\ }\href
  {https://doi.org/10.1103/PhysRevA.91.062107} {\bibfield  {journal} {\bibinfo
  {journal} {Phys. Rev. A}\ }\textbf {\bibinfo {volume} {91}},\ \bibinfo
  {pages} {062107} (\bibinfo {year} {2015})}\BibitemShut {NoStop}%
\bibitem [{\citenamefont {Vaidman}(2017)}]{vaidman2017weak}%
  \BibitemOpen
  \bibfield  {author} {\bibinfo {author} {\bibfnamefont {L.}~\bibnamefont
  {Vaidman}},\ }\href@noop {} {\bibfield  {journal} {\bibinfo  {journal}
  {Philosophical Transactions of the Royal Society A: Mathematical, Physical
  and Engineering Sciences}\ }\textbf {\bibinfo {volume} {375}},\ \bibinfo
  {pages} {20160395} (\bibinfo {year} {2017})}\BibitemShut {NoStop}%
\bibitem [{\citenamefont {Dixon}\ \emph {et~al.}(2009)\citenamefont {Dixon},
  \citenamefont {Starling}, \citenamefont {Jordan},\ and\ \citenamefont
  {Howell}}]{PhysRevLett.102.173601}%
  \BibitemOpen
  \bibfield  {author} {\bibinfo {author} {\bibfnamefont {P.~B.}\ \bibnamefont
  {Dixon}}, \bibinfo {author} {\bibfnamefont {D.~J.}\ \bibnamefont {Starling}},
  \bibinfo {author} {\bibfnamefont {A.~N.}\ \bibnamefont {Jordan}},\ and\
  \bibinfo {author} {\bibfnamefont {J.~C.}\ \bibnamefont {Howell}},\ }\href
  {https://doi.org/10.1103/PhysRevLett.102.173601} {\bibfield  {journal}
  {\bibinfo  {journal} {Phys. Rev. Lett.}\ }\textbf {\bibinfo {volume} {102}},\
  \bibinfo {pages} {173601} (\bibinfo {year} {2009})}\BibitemShut {NoStop}%
\bibitem [{\citenamefont {Ferrie}\ and\ \citenamefont
  {Combes}(2014)}]{PhysRevLett.112.040406}%
  \BibitemOpen
  \bibfield  {author} {\bibinfo {author} {\bibfnamefont {C.}~\bibnamefont
  {Ferrie}}\ and\ \bibinfo {author} {\bibfnamefont {J.}~\bibnamefont
  {Combes}},\ }\href {https://doi.org/10.1103/PhysRevLett.112.040406}
  {\bibfield  {journal} {\bibinfo  {journal} {Phys. Rev. Lett.}\ }\textbf
  {\bibinfo {volume} {112}},\ \bibinfo {pages} {040406} (\bibinfo {year}
  {2014})}\BibitemShut {NoStop}%
\bibitem [{\citenamefont {Combes}\ \emph {et~al.}(2014)\citenamefont {Combes},
  \citenamefont {Ferrie}, \citenamefont {Jiang},\ and\ \citenamefont
  {Caves}}]{combes2014quantum}%
  \BibitemOpen
  \bibfield  {author} {\bibinfo {author} {\bibfnamefont {J.}~\bibnamefont
  {Combes}}, \bibinfo {author} {\bibfnamefont {C.}~\bibnamefont {Ferrie}},
  \bibinfo {author} {\bibfnamefont {Z.}~\bibnamefont {Jiang}},\ and\ \bibinfo
  {author} {\bibfnamefont {C.~M.}\ \bibnamefont {Caves}},\ }\href@noop {}
  {\bibfield  {journal} {\bibinfo  {journal} {Physical Review A}\ }\textbf
  {\bibinfo {volume} {89}},\ \bibinfo {pages} {052117} (\bibinfo {year}
  {2014})}\BibitemShut {NoStop}%
\bibitem [{\citenamefont {Jordan}\ \emph {et~al.}(2014)\citenamefont {Jordan},
  \citenamefont {Martínez-Rincón},\ and\ \citenamefont
  {Howell}}]{2014Jordan}%
  \BibitemOpen
  \bibfield  {author} {\bibinfo {author} {\bibfnamefont {A.~N.}\ \bibnamefont
  {Jordan}}, \bibinfo {author} {\bibfnamefont {J.}~\bibnamefont
  {Martínez-Rincón}},\ and\ \bibinfo {author} {\bibfnamefont {J.~C.}\
  \bibnamefont {Howell}},\ }\bibfield  {journal} {\bibinfo  {journal} {Physical
  Review X}\ }\textbf {\bibinfo {volume} {4}},\ \href
  {https://doi.org/10.1103/physrevx.4.011031} {10.1103/physrevx.4.011031}
  (\bibinfo {year} {2014})\BibitemShut {NoStop}%
\bibitem [{\citenamefont {Knee}\ \emph {et~al.}(2016)\citenamefont {Knee},
  \citenamefont {Combes}, \citenamefont {Ferrie},\ and\ \citenamefont
  {Gauger}}]{KneeCombesFerrieGauger+2016}%
  \BibitemOpen
  \bibfield  {author} {\bibinfo {author} {\bibfnamefont {G.~C.}\ \bibnamefont
  {Knee}}, \bibinfo {author} {\bibfnamefont {J.}~\bibnamefont {Combes}},
  \bibinfo {author} {\bibfnamefont {C.}~\bibnamefont {Ferrie}},\ and\ \bibinfo
  {author} {\bibfnamefont {E.~M.}\ \bibnamefont {Gauger}},\ }\bibfield
  {journal} {\bibinfo  {journal} {Quantum Measurements and Quantum Metrology}\
  }\textbf {\bibinfo {volume} {3}},\ \href
  {https://doi.org/doi:10.1515/qmetro-2016-0006} {doi:10.1515/qmetro-2016-0006}
  (\bibinfo {year} {2016})\BibitemShut {NoStop}%
\bibitem [{\citenamefont {Vetrivelan}\ and\ \citenamefont
  {Vinjanampathy}(2022)}]{vetrivelan2022near}%
  \BibitemOpen
  \bibfield  {author} {\bibinfo {author} {\bibfnamefont {M.}~\bibnamefont
  {Vetrivelan}}\ and\ \bibinfo {author} {\bibfnamefont {S.}~\bibnamefont
  {Vinjanampathy}},\ }\href@noop {} {\bibfield  {journal} {\bibinfo  {journal}
  {Quantum Science and Technology}\ }\textbf {\bibinfo {volume} {7}},\ \bibinfo
  {pages} {025012} (\bibinfo {year} {2022})}\BibitemShut {NoStop}%
\bibitem [{\citenamefont {Jullien}\ \emph {et~al.}(2014)\citenamefont
  {Jullien}, \citenamefont {Roulleau}, \citenamefont {Roche}, \citenamefont
  {Cavanna}, \citenamefont {Jin},\ and\ \citenamefont
  {Glattli}}]{jullien2014quantum}%
  \BibitemOpen
  \bibfield  {author} {\bibinfo {author} {\bibfnamefont {T.}~\bibnamefont
  {Jullien}}, \bibinfo {author} {\bibfnamefont {P.}~\bibnamefont {Roulleau}},
  \bibinfo {author} {\bibfnamefont {B.}~\bibnamefont {Roche}}, \bibinfo
  {author} {\bibfnamefont {A.}~\bibnamefont {Cavanna}}, \bibinfo {author}
  {\bibfnamefont {Y.}~\bibnamefont {Jin}},\ and\ \bibinfo {author}
  {\bibfnamefont {D.}~\bibnamefont {Glattli}},\ }\href@noop {} {\bibfield
  {journal} {\bibinfo  {journal} {Nature}\ }\textbf {\bibinfo {volume} {514}},\
  \bibinfo {pages} {603} (\bibinfo {year} {2014})}\BibitemShut {NoStop}%
\bibitem [{\citenamefont {Samuelsson}\ and\ \citenamefont
  {B{\"u}ttiker}(2006)}]{samuelsson2006quantum}%
  \BibitemOpen
  \bibfield  {author} {\bibinfo {author} {\bibfnamefont {P.}~\bibnamefont
  {Samuelsson}}\ and\ \bibinfo {author} {\bibfnamefont {M.}~\bibnamefont
  {B{\"u}ttiker}},\ }\href@noop {} {\bibfield  {journal} {\bibinfo  {journal}
  {Physical Review B}\ }\textbf {\bibinfo {volume} {73}},\ \bibinfo {pages}
  {041305} (\bibinfo {year} {2006})}\BibitemShut {NoStop}%
\bibitem [{\citenamefont {Tam}\ \emph {et~al.}(2021)\citenamefont {Tam},
  \citenamefont {Flindt},\ and\ \citenamefont {Brange}}]{PhysRevB.104.245425}%
  \BibitemOpen
  \bibfield  {author} {\bibinfo {author} {\bibfnamefont {M.}~\bibnamefont
  {Tam}}, \bibinfo {author} {\bibfnamefont {C.}~\bibnamefont {Flindt}},\ and\
  \bibinfo {author} {\bibfnamefont {F.}~\bibnamefont {Brange}},\ }\href
  {https://doi.org/10.1103/PhysRevB.104.245425} {\bibfield  {journal} {\bibinfo
   {journal} {Phys. Rev. B}\ }\textbf {\bibinfo {volume} {104}},\ \bibinfo
  {pages} {245425} (\bibinfo {year} {2021})}\BibitemShut {NoStop}%
\bibitem [{\citenamefont {Ranni}\ \emph {et~al.}(2021)\citenamefont {Ranni},
  \citenamefont {Brange}, \citenamefont {Mannila}, \citenamefont {Flindt},\
  and\ \citenamefont {Maisi}}]{ranni2021real}%
  \BibitemOpen
  \bibfield  {author} {\bibinfo {author} {\bibfnamefont {A.}~\bibnamefont
  {Ranni}}, \bibinfo {author} {\bibfnamefont {F.}~\bibnamefont {Brange}},
  \bibinfo {author} {\bibfnamefont {E.~T.}\ \bibnamefont {Mannila}}, \bibinfo
  {author} {\bibfnamefont {C.}~\bibnamefont {Flindt}},\ and\ \bibinfo {author}
  {\bibfnamefont {V.~F.}\ \bibnamefont {Maisi}},\ }\href@noop {} {\bibfield
  {journal} {\bibinfo  {journal} {Nature communications}\ }\textbf {\bibinfo
  {volume} {12}},\ \bibinfo {pages} {1} (\bibinfo {year} {2021})}\BibitemShut
  {NoStop}%
\bibitem [{\citenamefont {Nigg}\ \emph {et~al.}(2015)\citenamefont {Nigg},
  \citenamefont {Tiwari}, \citenamefont {Walter},\ and\ \citenamefont
  {Schmidt}}]{PhysRevB.91.094516}%
  \BibitemOpen
  \bibfield  {author} {\bibinfo {author} {\bibfnamefont {S.~E.}\ \bibnamefont
  {Nigg}}, \bibinfo {author} {\bibfnamefont {R.~P.}\ \bibnamefont {Tiwari}},
  \bibinfo {author} {\bibfnamefont {S.}~\bibnamefont {Walter}},\ and\ \bibinfo
  {author} {\bibfnamefont {T.~L.}\ \bibnamefont {Schmidt}},\ }\href
  {https://doi.org/10.1103/PhysRevB.91.094516} {\bibfield  {journal} {\bibinfo
  {journal} {Phys. Rev. B}\ }\textbf {\bibinfo {volume} {91}},\ \bibinfo
  {pages} {094516} (\bibinfo {year} {2015})}\BibitemShut {NoStop}%
\bibitem [{\citenamefont {Brange}\ \emph {et~al.}(2021)\citenamefont {Brange},
  \citenamefont {Prech},\ and\ \citenamefont
  {Flindt}}]{PhysRevLett.127.237701}%
  \BibitemOpen
  \bibfield  {author} {\bibinfo {author} {\bibfnamefont {F.}~\bibnamefont
  {Brange}}, \bibinfo {author} {\bibfnamefont {K.}~\bibnamefont {Prech}},\ and\
  \bibinfo {author} {\bibfnamefont {C.}~\bibnamefont {Flindt}},\ }\href
  {https://doi.org/10.1103/PhysRevLett.127.237701} {\bibfield  {journal}
  {\bibinfo  {journal} {Phys. Rev. Lett.}\ }\textbf {\bibinfo {volume} {127}},\
  \bibinfo {pages} {237701} (\bibinfo {year} {2021})}\BibitemShut {NoStop}%
\bibitem [{\citenamefont {Deacon}\ \emph {et~al.}(2015)\citenamefont {Deacon},
  \citenamefont {Oiwa}, \citenamefont {Sailer}, \citenamefont {Baba},
  \citenamefont {Kanai}, \citenamefont {Shibata}, \citenamefont {Hirakawa},\
  and\ \citenamefont {Tarucha}}]{deacon2015cooper}%
  \BibitemOpen
  \bibfield  {author} {\bibinfo {author} {\bibfnamefont {R.}~\bibnamefont
  {Deacon}}, \bibinfo {author} {\bibfnamefont {A.}~\bibnamefont {Oiwa}},
  \bibinfo {author} {\bibfnamefont {J.}~\bibnamefont {Sailer}}, \bibinfo
  {author} {\bibfnamefont {S.}~\bibnamefont {Baba}}, \bibinfo {author}
  {\bibfnamefont {Y.}~\bibnamefont {Kanai}}, \bibinfo {author} {\bibfnamefont
  {K.}~\bibnamefont {Shibata}}, \bibinfo {author} {\bibfnamefont
  {K.}~\bibnamefont {Hirakawa}},\ and\ \bibinfo {author} {\bibfnamefont
  {S.}~\bibnamefont {Tarucha}},\ }\href@noop {} {\bibfield  {journal} {\bibinfo
   {journal} {Nature communications}\ }\textbf {\bibinfo {volume} {6}},\
  \bibinfo {pages} {1} (\bibinfo {year} {2015})}\BibitemShut {NoStop}%
\bibitem [{\citenamefont {Pfaff}\ \emph {et~al.}(2013)\citenamefont {Pfaff},
  \citenamefont {Taminiau}, \citenamefont {Robledo}, \citenamefont {Bernien},
  \citenamefont {Markham}, \citenamefont {Twitchen},\ and\ \citenamefont
  {Hanson}}]{pfaff2013demonstration}%
  \BibitemOpen
  \bibfield  {author} {\bibinfo {author} {\bibfnamefont {W.}~\bibnamefont
  {Pfaff}}, \bibinfo {author} {\bibfnamefont {T.~H.}\ \bibnamefont {Taminiau}},
  \bibinfo {author} {\bibfnamefont {L.}~\bibnamefont {Robledo}}, \bibinfo
  {author} {\bibfnamefont {H.}~\bibnamefont {Bernien}}, \bibinfo {author}
  {\bibfnamefont {M.}~\bibnamefont {Markham}}, \bibinfo {author} {\bibfnamefont
  {D.~J.}\ \bibnamefont {Twitchen}},\ and\ \bibinfo {author} {\bibfnamefont
  {R.}~\bibnamefont {Hanson}},\ }\href@noop {} {\bibfield  {journal} {\bibinfo
  {journal} {Nature Physics}\ }\textbf {\bibinfo {volume} {9}},\ \bibinfo
  {pages} {29} (\bibinfo {year} {2013})}\BibitemShut {NoStop}%
\bibitem [{\citenamefont {Ionicioiu}\ \emph {et~al.}(2001)\citenamefont
  {Ionicioiu}, \citenamefont {Zanardi},\ and\ \citenamefont
  {Rossi}}]{PhysRevA.63.050101}%
  \BibitemOpen
  \bibfield  {author} {\bibinfo {author} {\bibfnamefont {R.}~\bibnamefont
  {Ionicioiu}}, \bibinfo {author} {\bibfnamefont {P.}~\bibnamefont {Zanardi}},\
  and\ \bibinfo {author} {\bibfnamefont {F.}~\bibnamefont {Rossi}},\ }\href
  {https://doi.org/10.1103/PhysRevA.63.050101} {\bibfield  {journal} {\bibinfo
  {journal} {Phys. Rev. A}\ }\textbf {\bibinfo {volume} {63}},\ \bibinfo
  {pages} {050101} (\bibinfo {year} {2001})}\BibitemShut {NoStop}%
\bibitem [{\citenamefont {Bednorz}\ and\ \citenamefont
  {Belzig}(2011)}]{PhysRevB.83.125304}%
  \BibitemOpen
  \bibfield  {author} {\bibinfo {author} {\bibfnamefont {A.}~\bibnamefont
  {Bednorz}}\ and\ \bibinfo {author} {\bibfnamefont {W.}~\bibnamefont
  {Belzig}},\ }\href {https://doi.org/10.1103/PhysRevB.83.125304} {\bibfield
  {journal} {\bibinfo  {journal} {Phys. Rev. B}\ }\textbf {\bibinfo {volume}
  {83}},\ \bibinfo {pages} {125304} (\bibinfo {year} {2011})}\BibitemShut
  {NoStop}%
\bibitem [{\citenamefont {Zhou}\ \emph {et~al.}(2019)\citenamefont {Zhou},
  \citenamefont {Taguchi}, \citenamefont {Kawaguchi}, \citenamefont {Tanaka},\
  and\ \citenamefont {Law}}]{zhou2019spin}%
  \BibitemOpen
  \bibfield  {author} {\bibinfo {author} {\bibfnamefont {B.~T.}\ \bibnamefont
  {Zhou}}, \bibinfo {author} {\bibfnamefont {K.}~\bibnamefont {Taguchi}},
  \bibinfo {author} {\bibfnamefont {Y.}~\bibnamefont {Kawaguchi}}, \bibinfo
  {author} {\bibfnamefont {Y.}~\bibnamefont {Tanaka}},\ and\ \bibinfo {author}
  {\bibfnamefont {K.~T.}\ \bibnamefont {Law}},\ }\href@noop {} {\bibfield
  {journal} {\bibinfo  {journal} {Communications Physics}\ }\textbf {\bibinfo
  {volume} {2}},\ \bibinfo {pages} {1} (\bibinfo {year} {2019})}\BibitemShut
  {NoStop}%
\bibitem [{\citenamefont {Zhang}\ \emph {et~al.}(2020)\citenamefont {Zhang},
  \citenamefont {Hou}, \citenamefont {Zhao}, \citenamefont {Guo}, \citenamefont
  {Liu}, \citenamefont {Li}, \citenamefont {Ren}, \citenamefont {Sun},\ and\
  \citenamefont {He}}]{PhysRevB.102.081403}%
  \BibitemOpen
  \bibfield  {author} {\bibinfo {author} {\bibfnamefont {Y.}~\bibnamefont
  {Zhang}}, \bibinfo {author} {\bibfnamefont {Z.}~\bibnamefont {Hou}}, \bibinfo
  {author} {\bibfnamefont {Y.-X.}\ \bibnamefont {Zhao}}, \bibinfo {author}
  {\bibfnamefont {Z.-H.}\ \bibnamefont {Guo}}, \bibinfo {author} {\bibfnamefont
  {Y.-W.}\ \bibnamefont {Liu}}, \bibinfo {author} {\bibfnamefont {S.-Y.}\
  \bibnamefont {Li}}, \bibinfo {author} {\bibfnamefont {Y.-N.}\ \bibnamefont
  {Ren}}, \bibinfo {author} {\bibfnamefont {Q.-F.}\ \bibnamefont {Sun}},\ and\
  \bibinfo {author} {\bibfnamefont {L.}~\bibnamefont {He}},\ }\href
  {https://doi.org/10.1103/PhysRevB.102.081403} {\bibfield  {journal} {\bibinfo
   {journal} {Phys. Rev. B}\ }\textbf {\bibinfo {volume} {102}},\ \bibinfo
  {pages} {081403} (\bibinfo {year} {2020})}\BibitemShut {NoStop}%
\bibitem [{\citenamefont {Li}\ \emph {et~al.}(2014)\citenamefont {Li},
  \citenamefont {Ludwig}, \citenamefont {Low}, \citenamefont {Chernikov},
  \citenamefont {Cui}, \citenamefont {Arefe}, \citenamefont {Kim},
  \citenamefont {van~der Zande}, \citenamefont {Rigosi}, \citenamefont {Hill},
  \citenamefont {Kim}, \citenamefont {Hone}, \citenamefont {Li}, \citenamefont
  {Smirnov},\ and\ \citenamefont {Heinz}}]{PhysRevLett.113.266804}%
  \BibitemOpen
  \bibfield  {author} {\bibinfo {author} {\bibfnamefont {Y.}~\bibnamefont
  {Li}}, \bibinfo {author} {\bibfnamefont {J.}~\bibnamefont {Ludwig}}, \bibinfo
  {author} {\bibfnamefont {T.}~\bibnamefont {Low}}, \bibinfo {author}
  {\bibfnamefont {A.}~\bibnamefont {Chernikov}}, \bibinfo {author}
  {\bibfnamefont {X.}~\bibnamefont {Cui}}, \bibinfo {author} {\bibfnamefont
  {G.}~\bibnamefont {Arefe}}, \bibinfo {author} {\bibfnamefont {Y.~D.}\
  \bibnamefont {Kim}}, \bibinfo {author} {\bibfnamefont {A.~M.}\ \bibnamefont
  {van~der Zande}}, \bibinfo {author} {\bibfnamefont {A.}~\bibnamefont
  {Rigosi}}, \bibinfo {author} {\bibfnamefont {H.~M.}\ \bibnamefont {Hill}},
  \bibinfo {author} {\bibfnamefont {S.~H.}\ \bibnamefont {Kim}}, \bibinfo
  {author} {\bibfnamefont {J.}~\bibnamefont {Hone}}, \bibinfo {author}
  {\bibfnamefont {Z.}~\bibnamefont {Li}}, \bibinfo {author} {\bibfnamefont
  {D.}~\bibnamefont {Smirnov}},\ and\ \bibinfo {author} {\bibfnamefont {T.~F.}\
  \bibnamefont {Heinz}},\ }\href
  {https://doi.org/10.1103/PhysRevLett.113.266804} {\bibfield  {journal}
  {\bibinfo  {journal} {Phys. Rev. Lett.}\ }\textbf {\bibinfo {volume} {113}},\
  \bibinfo {pages} {266804} (\bibinfo {year} {2014})}\BibitemShut {NoStop}%
\bibitem [{\citenamefont {Zhang}\ \emph {et~al.}(2019)\citenamefont {Zhang},
  \citenamefont {Lai}, \citenamefont {Dohner}, \citenamefont {Moon},
  \citenamefont {Taniguchi}, \citenamefont {Watanabe}, \citenamefont
  {Smirnov},\ and\ \citenamefont {Heinz}}]{PhysRevLett.122.127401}%
  \BibitemOpen
  \bibfield  {author} {\bibinfo {author} {\bibfnamefont {X.-X.}\ \bibnamefont
  {Zhang}}, \bibinfo {author} {\bibfnamefont {Y.}~\bibnamefont {Lai}}, \bibinfo
  {author} {\bibfnamefont {E.}~\bibnamefont {Dohner}}, \bibinfo {author}
  {\bibfnamefont {S.}~\bibnamefont {Moon}}, \bibinfo {author} {\bibfnamefont
  {T.}~\bibnamefont {Taniguchi}}, \bibinfo {author} {\bibfnamefont
  {K.}~\bibnamefont {Watanabe}}, \bibinfo {author} {\bibfnamefont
  {D.}~\bibnamefont {Smirnov}},\ and\ \bibinfo {author} {\bibfnamefont {T.~F.}\
  \bibnamefont {Heinz}},\ }\href
  {https://doi.org/10.1103/PhysRevLett.122.127401} {\bibfield  {journal}
  {\bibinfo  {journal} {Phys. Rev. Lett.}\ }\textbf {\bibinfo {volume} {122}},\
  \bibinfo {pages} {127401} (\bibinfo {year} {2019})}\BibitemShut {NoStop}%
\bibitem [{\citenamefont {Dankert}\ and\ \citenamefont
  {Dash}(2017)}]{dankert2017electrical}%
  \BibitemOpen
  \bibfield  {author} {\bibinfo {author} {\bibfnamefont {A.}~\bibnamefont
  {Dankert}}\ and\ \bibinfo {author} {\bibfnamefont {S.~P.}\ \bibnamefont
  {Dash}},\ }\href@noop {} {\bibfield  {journal} {\bibinfo  {journal} {Nature
  communications}\ }\textbf {\bibinfo {volume} {8}},\ \bibinfo {pages} {1}
  (\bibinfo {year} {2017})}\BibitemShut {NoStop}%
\bibitem [{\citenamefont {Khokhriakov}\ \emph {et~al.}(2020)\citenamefont
  {Khokhriakov}, \citenamefont {Hoque}, \citenamefont {Karpiak},\ and\
  \citenamefont {Dash}}]{khokhriakov2020gate}%
  \BibitemOpen
  \bibfield  {author} {\bibinfo {author} {\bibfnamefont {D.}~\bibnamefont
  {Khokhriakov}}, \bibinfo {author} {\bibfnamefont {A.~M.}\ \bibnamefont
  {Hoque}}, \bibinfo {author} {\bibfnamefont {B.}~\bibnamefont {Karpiak}},\
  and\ \bibinfo {author} {\bibfnamefont {S.~P.}\ \bibnamefont {Dash}},\
  }\href@noop {} {\bibfield  {journal} {\bibinfo  {journal} {Nature
  communications}\ }\textbf {\bibinfo {volume} {11}},\ \bibinfo {pages} {1}
  (\bibinfo {year} {2020})}\BibitemShut {NoStop}%
\bibitem [{\citenamefont {Kamalakar}\ \emph {et~al.}(2016)\citenamefont
  {Kamalakar}, \citenamefont {Dankert}, \citenamefont {Kelly},\ and\
  \citenamefont {Dash}}]{kamalakar2016inversion}%
  \BibitemOpen
  \bibfield  {author} {\bibinfo {author} {\bibfnamefont {M.~V.}\ \bibnamefont
  {Kamalakar}}, \bibinfo {author} {\bibfnamefont {A.}~\bibnamefont {Dankert}},
  \bibinfo {author} {\bibfnamefont {P.~J.}\ \bibnamefont {Kelly}},\ and\
  \bibinfo {author} {\bibfnamefont {S.~P.}\ \bibnamefont {Dash}},\ }\href@noop
  {} {\bibfield  {journal} {\bibinfo  {journal} {Scientific reports}\ }\textbf
  {\bibinfo {volume} {6}},\ \bibinfo {pages} {1} (\bibinfo {year}
  {2016})}\BibitemShut {NoStop}%
\bibitem [{\citenamefont {Tsitsishvili}\ \emph {et~al.}(2004)\citenamefont
  {Tsitsishvili}, \citenamefont {Lozano},\ and\ \citenamefont
  {Gogolin}}]{PhysRevB.70.115316}%
  \BibitemOpen
  \bibfield  {author} {\bibinfo {author} {\bibfnamefont {E.}~\bibnamefont
  {Tsitsishvili}}, \bibinfo {author} {\bibfnamefont {G.~S.}\ \bibnamefont
  {Lozano}},\ and\ \bibinfo {author} {\bibfnamefont {A.~O.}\ \bibnamefont
  {Gogolin}},\ }\href {https://doi.org/10.1103/PhysRevB.70.115316} {\bibfield
  {journal} {\bibinfo  {journal} {Phys. Rev. B}\ }\textbf {\bibinfo {volume}
  {70}},\ \bibinfo {pages} {115316} (\bibinfo {year} {2004})}\BibitemShut
  {NoStop}%
\bibitem [{\citenamefont {S{\'a}nchez}\ \emph {et~al.}(2013)\citenamefont
  {S{\'a}nchez}, \citenamefont {Vila}, \citenamefont {Desfonds}, \citenamefont
  {Gambarelli}, \citenamefont {Attan{\'e}}, \citenamefont {De~Teresa},
  \citenamefont {Mag{\'e}n},\ and\ \citenamefont {Fert}}]{sanchez2013spin}%
  \BibitemOpen
  \bibfield  {author} {\bibinfo {author} {\bibfnamefont {J.}~\bibnamefont
  {S{\'a}nchez}}, \bibinfo {author} {\bibfnamefont {L.}~\bibnamefont {Vila}},
  \bibinfo {author} {\bibfnamefont {G.}~\bibnamefont {Desfonds}}, \bibinfo
  {author} {\bibfnamefont {S.}~\bibnamefont {Gambarelli}}, \bibinfo {author}
  {\bibfnamefont {J.}~\bibnamefont {Attan{\'e}}}, \bibinfo {author}
  {\bibfnamefont {J.}~\bibnamefont {De~Teresa}}, \bibinfo {author}
  {\bibfnamefont {C.}~\bibnamefont {Mag{\'e}n}},\ and\ \bibinfo {author}
  {\bibfnamefont {A.}~\bibnamefont {Fert}},\ }\href@noop {} {\bibfield
  {journal} {\bibinfo  {journal} {Nature communications}\ }\textbf {\bibinfo
  {volume} {4}},\ \bibinfo {pages} {1} (\bibinfo {year} {2013})}\BibitemShut
  {NoStop}%
\bibitem [{\citenamefont {Steinberg}(1995)}]{Steinberg1995}%
  \BibitemOpen
  \bibfield  {author} {\bibinfo {author} {\bibfnamefont {A.~M.}\ \bibnamefont
  {Steinberg}},\ }\href {https://doi.org/10.1103/physrevlett.74.2405}
  {\bibfield  {journal} {\bibinfo  {journal} {Physical Review Letters}\
  }\textbf {\bibinfo {volume} {74}},\ \bibinfo {pages} {2405–2409} (\bibinfo
  {year} {1995})}\BibitemShut {NoStop}%
\bibitem [{\citenamefont {Mathew}\ \emph {et~al.}(2022)\citenamefont {Mathew},
  \citenamefont {Camsari},\ and\ \citenamefont
  {Muralidharan}}]{PhysRevB.105.144418}%
  \BibitemOpen
  \bibfield  {author} {\bibinfo {author} {\bibfnamefont {A.}~\bibnamefont
  {Mathew}}, \bibinfo {author} {\bibfnamefont {K.~Y.}\ \bibnamefont
  {Camsari}},\ and\ \bibinfo {author} {\bibfnamefont {B.}~\bibnamefont
  {Muralidharan}},\ }\href {https://doi.org/10.1103/PhysRevB.105.144418}
  {\bibfield  {journal} {\bibinfo  {journal} {Phys. Rev. B}\ }\textbf {\bibinfo
  {volume} {105}},\ \bibinfo {pages} {144418} (\bibinfo {year}
  {2022})}\BibitemShut {NoStop}%
\bibitem [{\citenamefont {Ricco}\ and\ \citenamefont
  {Azbel}(1984)}]{PhysRevB.29.1970}%
  \BibitemOpen
  \bibfield  {author} {\bibinfo {author} {\bibfnamefont {B.}~\bibnamefont
  {Ricco}}\ and\ \bibinfo {author} {\bibfnamefont {M.~Y.}\ \bibnamefont
  {Azbel}},\ }\href {https://doi.org/10.1103/PhysRevB.29.1970} {\bibfield
  {journal} {\bibinfo  {journal} {Phys. Rev. B}\ }\textbf {\bibinfo {volume}
  {29}},\ \bibinfo {pages} {1970} (\bibinfo {year} {1984})}\BibitemShut
  {NoStop}%
\bibitem [{\citenamefont {Sun}\ \emph {et~al.}(1998)\citenamefont {Sun},
  \citenamefont {Haddad}, \citenamefont {Mazumder},\ and\ \citenamefont
  {Schulman}}]{sun1998resonant}%
  \BibitemOpen
  \bibfield  {author} {\bibinfo {author} {\bibfnamefont {J.~P.}\ \bibnamefont
  {Sun}}, \bibinfo {author} {\bibfnamefont {G.~I.}\ \bibnamefont {Haddad}},
  \bibinfo {author} {\bibfnamefont {P.}~\bibnamefont {Mazumder}},\ and\
  \bibinfo {author} {\bibfnamefont {J.~N.}\ \bibnamefont {Schulman}},\
  }\href@noop {} {\bibfield  {journal} {\bibinfo  {journal} {Proceedings of the
  IEEE}\ }\textbf {\bibinfo {volume} {86}},\ \bibinfo {pages} {641} (\bibinfo
  {year} {1998})}\BibitemShut {NoStop}%
\bibitem [{\citenamefont {Bj{\"o}rk}\ \emph {et~al.}(2002)\citenamefont
  {Bj{\"o}rk}, \citenamefont {Ohlsson}, \citenamefont {Thelander},
  \citenamefont {Persson}, \citenamefont {Deppert}, \citenamefont
  {Wallenberg},\ and\ \citenamefont {Samuelson}}]{bjork2002nanowire}%
  \BibitemOpen
  \bibfield  {author} {\bibinfo {author} {\bibfnamefont {M.}~\bibnamefont
  {Bj{\"o}rk}}, \bibinfo {author} {\bibfnamefont {B.}~\bibnamefont {Ohlsson}},
  \bibinfo {author} {\bibfnamefont {C.}~\bibnamefont {Thelander}}, \bibinfo
  {author} {\bibfnamefont {A.}~\bibnamefont {Persson}}, \bibinfo {author}
  {\bibfnamefont {K.}~\bibnamefont {Deppert}}, \bibinfo {author} {\bibfnamefont
  {L.}~\bibnamefont {Wallenberg}},\ and\ \bibinfo {author} {\bibfnamefont
  {L.}~\bibnamefont {Samuelson}},\ }\href@noop {} {\bibfield  {journal}
  {\bibinfo  {journal} {Applied Physics Letters}\ }\textbf {\bibinfo {volume}
  {81}},\ \bibinfo {pages} {4458} (\bibinfo {year} {2002})}\BibitemShut
  {NoStop}%
\bibitem [{\citenamefont {Mazumder}\ \emph {et~al.}(1998)\citenamefont
  {Mazumder}, \citenamefont {Kulkarni}, \citenamefont {Bhattacharya},
  \citenamefont {Sun},\ and\ \citenamefont {Haddad}}]{663544}%
  \BibitemOpen
  \bibfield  {author} {\bibinfo {author} {\bibfnamefont {P.}~\bibnamefont
  {Mazumder}}, \bibinfo {author} {\bibfnamefont {S.}~\bibnamefont {Kulkarni}},
  \bibinfo {author} {\bibfnamefont {M.}~\bibnamefont {Bhattacharya}}, \bibinfo
  {author} {\bibfnamefont {J.~P.}\ \bibnamefont {Sun}},\ and\ \bibinfo {author}
  {\bibfnamefont {G.}~\bibnamefont {Haddad}},\ }\href
  {https://doi.org/10.1109/5.663544} {\bibfield  {journal} {\bibinfo  {journal}
  {Proceedings of the IEEE}\ }\textbf {\bibinfo {volume} {86}},\ \bibinfo
  {pages} {664} (\bibinfo {year} {1998})}\BibitemShut {NoStop}%
\bibitem [{\citenamefont {Danielewicz}(1984)}]{DANIELEWICZ1984239}%
  \BibitemOpen
  \bibfield  {author} {\bibinfo {author} {\bibfnamefont {P.}~\bibnamefont
  {Danielewicz}},\ }\href
  {https://doi.org/https://doi.org/10.1016/0003-4916(84)90092-7} {\bibfield
  {journal} {\bibinfo  {journal} {Annals of Physics}\ }\textbf {\bibinfo
  {volume} {152}},\ \bibinfo {pages} {239 } (\bibinfo {year}
  {1984})}\BibitemShut {NoStop}%
\bibitem [{\citenamefont {Datta}(1997)}]{datta1997electronic}%
  \BibitemOpen
  \bibfield  {author} {\bibinfo {author} {\bibfnamefont {S.}~\bibnamefont
  {Datta}},\ }\href@noop {} {\emph {\bibinfo {title} {Electronic transport in
  mesoscopic systems}}}\ (\bibinfo  {publisher} {Cambridge university press},\
  \bibinfo {year} {1997})\BibitemShut {NoStop}%
\bibitem [{\citenamefont {Datta}(2005)}]{book91179691}%
  \BibitemOpen
  \bibfield  {author} {\bibinfo {author} {\bibfnamefont {S.}~\bibnamefont
  {Datta}},\ }\href@noop {} {\emph {\bibinfo {title} {Quantum Transport: Atom
  to Transistor}}},\ \bibinfo {edition} {2nd}\ ed.\ (\bibinfo  {publisher}
  {Cambridge University Press},\ \bibinfo {year} {2005})\BibitemShut {NoStop}%
\bibitem [{\citenamefont {Golizadeh-Mojarad}\ and\ \citenamefont
  {Datta}(2007)}]{PhysRevB.75.081301}%
  \BibitemOpen
  \bibfield  {author} {\bibinfo {author} {\bibfnamefont {R.}~\bibnamefont
  {Golizadeh-Mojarad}}\ and\ \bibinfo {author} {\bibfnamefont {S.}~\bibnamefont
  {Datta}},\ }\href {https://doi.org/10.1103/PhysRevB.75.081301} {\bibfield
  {journal} {\bibinfo  {journal} {Phys. Rev. B}\ }\textbf {\bibinfo {volume}
  {75}},\ \bibinfo {pages} {081301} (\bibinfo {year} {2007})}\BibitemShut
  {NoStop}%
\bibitem [{\citenamefont {{Sharma}}\ \emph {et~al.}(2016)\citenamefont
  {{Sharma}}, \citenamefont {{Tulapurkar}},\ and\ \citenamefont
  {{Muralidharan}}}]{7571106}%
  \BibitemOpen
  \bibfield  {author} {\bibinfo {author} {\bibfnamefont {A.}~\bibnamefont
  {{Sharma}}}, \bibinfo {author} {\bibfnamefont {A.}~\bibnamefont
  {{Tulapurkar}}},\ and\ \bibinfo {author} {\bibfnamefont {B.}~\bibnamefont
  {{Muralidharan}}},\ }\href {https://doi.org/10.1109/TED.2016.2606354}
  {\bibfield  {journal} {\bibinfo  {journal} {IEEE Transactions on Electron
  Devices}\ }\textbf {\bibinfo {volume} {63}},\ \bibinfo {pages} {4527}
  (\bibinfo {year} {2016})}\BibitemShut {NoStop}%
\bibitem [{\citenamefont {Sharma}\ \emph {et~al.}(2018)\citenamefont {Sharma},
  \citenamefont {Tulapurkar},\ and\ \citenamefont
  {Muralidharan}}]{doi:10.1063/1.5023159}%
  \BibitemOpen
  \bibfield  {author} {\bibinfo {author} {\bibfnamefont {A.}~\bibnamefont
  {Sharma}}, \bibinfo {author} {\bibfnamefont {A.~A.}\ \bibnamefont
  {Tulapurkar}},\ and\ \bibinfo {author} {\bibfnamefont {B.}~\bibnamefont
  {Muralidharan}},\ }\href {https://doi.org/10.1063/1.5023159} {\bibfield
  {journal} {\bibinfo  {journal} {Applied Physics Letters}\ }\textbf {\bibinfo
  {volume} {112}},\ \bibinfo {pages} {192404} (\bibinfo {year}
  {2018})}\BibitemShut {NoStop}%
\bibitem [{\citenamefont {Singha}\ and\ \citenamefont
  {Muralidharan}(2018)}]{doi:10.1063/1.5044254}%
  \BibitemOpen
  \bibfield  {author} {\bibinfo {author} {\bibfnamefont {A.}~\bibnamefont
  {Singha}}\ and\ \bibinfo {author} {\bibfnamefont {B.}~\bibnamefont
  {Muralidharan}},\ }\href@noop {} {\bibfield  {journal} {\bibinfo  {journal}
  {Journal of Applied Physics}\ }\textbf {\bibinfo {volume} {124}},\ \bibinfo
  {pages} {144901} (\bibinfo {year} {2018})}\BibitemShut {NoStop}%
\bibitem [{\citenamefont {Sharma}\ \emph {et~al.}(2017)\citenamefont {Sharma},
  \citenamefont {Tulapurkar},\ and\ \citenamefont
  {Muralidharan}}]{PhysRevApplied.8.064014}%
  \BibitemOpen
  \bibfield  {author} {\bibinfo {author} {\bibfnamefont {A.}~\bibnamefont
  {Sharma}}, \bibinfo {author} {\bibfnamefont {A.~A.}\ \bibnamefont
  {Tulapurkar}},\ and\ \bibinfo {author} {\bibfnamefont {B.}~\bibnamefont
  {Muralidharan}},\ }\href {https://doi.org/10.1103/PhysRevApplied.8.064014}
  {\bibfield  {journal} {\bibinfo  {journal} {Phys. Rev. Applied}\ }\textbf
  {\bibinfo {volume} {8}},\ \bibinfo {pages} {064014} (\bibinfo {year}
  {2017})}\BibitemShut {NoStop}%
\bibitem [{\citenamefont {Camsari}\ \emph {et~al.}(2020)\citenamefont
  {Camsari}, \citenamefont {Chowdhury},\ and\ \citenamefont
  {Datta}}]{camsari2020}%
  \BibitemOpen
  \bibfield  {author} {\bibinfo {author} {\bibfnamefont {K.~Y.}\ \bibnamefont
  {Camsari}}, \bibinfo {author} {\bibfnamefont {S.}~\bibnamefont {Chowdhury}},\
  and\ \bibinfo {author} {\bibfnamefont {S.}~\bibnamefont {Datta}},\
  }\href@noop {} {\bibinfo {title} {The non-equilibrium green function (negf)
  method}} (\bibinfo {year} {2020}),\ \Eprint
  {https://arxiv.org/abs/2008.01275} {arXiv:2008.01275 [cond-mat.mes-hall]}
  \BibitemShut {NoStop}%
\bibitem [{\citenamefont {Duse}\ \emph {et~al.}(2021)\citenamefont {Duse},
  \citenamefont {Sriram}, \citenamefont {Gharavi}, \citenamefont {Baugh},\ and\
  \citenamefont {Muralidharan}}]{Duse_2021}%
  \BibitemOpen
  \bibfield  {author} {\bibinfo {author} {\bibfnamefont {C.}~\bibnamefont
  {Duse}}, \bibinfo {author} {\bibfnamefont {P.}~\bibnamefont {Sriram}},
  \bibinfo {author} {\bibfnamefont {K.}~\bibnamefont {Gharavi}}, \bibinfo
  {author} {\bibfnamefont {J.}~\bibnamefont {Baugh}},\ and\ \bibinfo {author}
  {\bibfnamefont {B.}~\bibnamefont {Muralidharan}},\ }\href
  {https://doi.org/10.1088/1361-648x/ac0d16} {\bibfield  {journal} {\bibinfo
  {journal} {Journal of Physics: Condensed Matter}\ }\textbf {\bibinfo {volume}
  {33}},\ \bibinfo {pages} {365301} (\bibinfo {year} {2021})}\BibitemShut
  {NoStop}%
\bibitem [{\citenamefont {Aharonov}\ \emph {et~al.}(1988)\citenamefont
  {Aharonov}, \citenamefont {Albert},\ and\ \citenamefont
  {Vaidman}}]{PhysRevLett.60.1351}%
  \BibitemOpen
  \bibfield  {author} {\bibinfo {author} {\bibfnamefont {Y.}~\bibnamefont
  {Aharonov}}, \bibinfo {author} {\bibfnamefont {D.~Z.}\ \bibnamefont
  {Albert}},\ and\ \bibinfo {author} {\bibfnamefont {L.}~\bibnamefont
  {Vaidman}},\ }\href {https://doi.org/10.1103/PhysRevLett.60.1351} {\bibfield
  {journal} {\bibinfo  {journal} {Phys. Rev. Lett.}\ }\textbf {\bibinfo
  {volume} {60}},\ \bibinfo {pages} {1351} (\bibinfo {year}
  {1988})}\BibitemShut {NoStop}%
\bibitem [{\citenamefont {Duck}\ \emph {et~al.}(1989)\citenamefont {Duck},
  \citenamefont {Stevenson},\ and\ \citenamefont
  {Sudarshan}}]{PhysRevD.40.2112}%
  \BibitemOpen
  \bibfield  {author} {\bibinfo {author} {\bibfnamefont {I.~M.}\ \bibnamefont
  {Duck}}, \bibinfo {author} {\bibfnamefont {P.~M.}\ \bibnamefont
  {Stevenson}},\ and\ \bibinfo {author} {\bibfnamefont {E.~C.~G.}\ \bibnamefont
  {Sudarshan}},\ }\href {https://doi.org/10.1103/PhysRevD.40.2112} {\bibfield
  {journal} {\bibinfo  {journal} {Phys. Rev. D}\ }\textbf {\bibinfo {volume}
  {40}},\ \bibinfo {pages} {2112} (\bibinfo {year} {1989})}\BibitemShut
  {NoStop}%
\bibitem [{\citenamefont {Leggett}(1989)}]{PhysRevLett.62.2325}%
  \BibitemOpen
  \bibfield  {author} {\bibinfo {author} {\bibfnamefont {A.~J.}\ \bibnamefont
  {Leggett}},\ }\href {https://doi.org/10.1103/PhysRevLett.62.2325} {\bibfield
  {journal} {\bibinfo  {journal} {Phys. Rev. Lett.}\ }\textbf {\bibinfo
  {volume} {62}},\ \bibinfo {pages} {2325} (\bibinfo {year}
  {1989})}\BibitemShut {NoStop}%
\bibitem [{\citenamefont {Dressel}\ \emph {et~al.}(2014)\citenamefont
  {Dressel}, \citenamefont {Malik}, \citenamefont {Miatto}, \citenamefont
  {Jordan},\ and\ \citenamefont {Boyd}}]{Dressel2014}%
  \BibitemOpen
  \bibfield  {author} {\bibinfo {author} {\bibfnamefont {J.}~\bibnamefont
  {Dressel}}, \bibinfo {author} {\bibfnamefont {M.}~\bibnamefont {Malik}},
  \bibinfo {author} {\bibfnamefont {F.}~\bibnamefont {Miatto}}, \bibinfo
  {author} {\bibfnamefont {A.}~\bibnamefont {Jordan}},\ and\ \bibinfo {author}
  {\bibfnamefont {R.}~\bibnamefont {Boyd}},\ }\href
  {https://doi.org/10.1103/RevModPhys.86.307} {\bibfield  {journal} {\bibinfo
  {journal} {Review of Modern Physics}\ }\textbf {\bibinfo {volume} {86}},\
  \bibinfo {pages} {307} (\bibinfo {year} {2014})}\BibitemShut {NoStop}%
\bibitem [{\citenamefont {Meir}\ and\ \citenamefont
  {Wingreen}(1992)}]{Meir-Wingreen-1992}%
  \BibitemOpen
  \bibfield  {author} {\bibinfo {author} {\bibfnamefont {Y.}~\bibnamefont
  {Meir}}\ and\ \bibinfo {author} {\bibfnamefont {N.~S.}\ \bibnamefont
  {Wingreen}},\ }\href {https://doi.org/10.1103/PhysRevLett.68.2512} {\bibfield
   {journal} {\bibinfo  {journal} {Phys. Rev. Lett.}\ }\textbf {\bibinfo
  {volume} {68}},\ \bibinfo {pages} {2512} (\bibinfo {year}
  {1992})}\BibitemShut {NoStop}%
\bibitem [{\citenamefont {Haug}\ and\ \citenamefont
  {Jauho}(2007)}]{haug2007quantum}%
  \BibitemOpen
  \bibfield  {author} {\bibinfo {author} {\bibfnamefont {H.}~\bibnamefont
  {Haug}}\ and\ \bibinfo {author} {\bibfnamefont {A.}~\bibnamefont {Jauho}},\
  }\href@noop {} {\emph {\bibinfo {title} {Quantum Kinetics in Transport and
  Optics of Semiconductors}}},\ Springer Series in Solid-State Sciences\
  (\bibinfo  {publisher} {Springer Berlin Heidelberg},\ \bibinfo {year}
  {2007})\BibitemShut {NoStop}%
\bibitem [{\citenamefont {Zatelli}\ \emph {et~al.}(2020)\citenamefont
  {Zatelli}, \citenamefont {Benedetti},\ and\ \citenamefont
  {Paris}}]{e22111321}%
  \BibitemOpen
  \bibfield  {author} {\bibinfo {author} {\bibfnamefont {F.}~\bibnamefont
  {Zatelli}}, \bibinfo {author} {\bibfnamefont {C.}~\bibnamefont {Benedetti}},\
  and\ \bibinfo {author} {\bibfnamefont {M.~G.~A.}\ \bibnamefont {Paris}},\
  }\bibfield  {journal} {\bibinfo  {journal} {Entropy}\ }\textbf {\bibinfo
  {volume} {22}},\ \href {https://doi.org/10.3390/e22111321}
  {10.3390/e22111321} (\bibinfo {year} {2020})\BibitemShut {NoStop}%
\bibitem [{\citenamefont {Sakurai}\ and\ \citenamefont
  {Napolitano}(2017)}]{sakurai_napolitano_2017}%
  \BibitemOpen
  \bibfield  {author} {\bibinfo {author} {\bibfnamefont {J.~J.}\ \bibnamefont
  {Sakurai}}\ and\ \bibinfo {author} {\bibfnamefont {J.}~\bibnamefont
  {Napolitano}},\ }\href {https://doi.org/10.1017/9781108499996} {\emph
  {\bibinfo {title} {Modern Quantum Mechanics}}},\ \bibinfo {edition} {2nd}\
  ed.\ (\bibinfo  {publisher} {Cambridge University Press},\ \bibinfo {year}
  {2017})\BibitemShut {NoStop}%
\bibitem [{\citenamefont {Liu}\ \emph {et~al.}(2016)\citenamefont {Liu},
  \citenamefont {Chen}, \citenamefont {Jing},\ and\ \citenamefont
  {Wang}}]{Liu_2016}%
  \BibitemOpen
  \bibfield  {author} {\bibinfo {author} {\bibfnamefont {J.}~\bibnamefont
  {Liu}}, \bibinfo {author} {\bibfnamefont {J.}~\bibnamefont {Chen}}, \bibinfo
  {author} {\bibfnamefont {X.-X.}\ \bibnamefont {Jing}},\ and\ \bibinfo
  {author} {\bibfnamefont {X.}~\bibnamefont {Wang}},\ }\href
  {https://doi.org/10.1088/1751-8113/49/27/275302} {\bibfield  {journal}
  {\bibinfo  {journal} {Journal of Physics A: Mathematical and Theoretical}\
  }\textbf {\bibinfo {volume} {49}},\ \bibinfo {pages} {275302} (\bibinfo
  {year} {2016})}\BibitemShut {NoStop}%
\bibitem [{\citenamefont {Boixo}\ \emph {et~al.}(2007)\citenamefont {Boixo},
  \citenamefont {Flammia}, \citenamefont {Caves},\ and\ \citenamefont
  {Geremia}}]{PhysRevLett.98.090401}%
  \BibitemOpen
  \bibfield  {author} {\bibinfo {author} {\bibfnamefont {S.}~\bibnamefont
  {Boixo}}, \bibinfo {author} {\bibfnamefont {S.~T.}\ \bibnamefont {Flammia}},
  \bibinfo {author} {\bibfnamefont {C.~M.}\ \bibnamefont {Caves}},\ and\
  \bibinfo {author} {\bibfnamefont {J.}~\bibnamefont {Geremia}},\ }\href
  {https://doi.org/10.1103/PhysRevLett.98.090401} {\bibfield  {journal}
  {\bibinfo  {journal} {Phys. Rev. Lett.}\ }\textbf {\bibinfo {volume} {98}},\
  \bibinfo {pages} {090401} (\bibinfo {year} {2007})}\BibitemShut {NoStop}%
\bibitem [{\citenamefont {Braunstein}\ \emph {et~al.}(1996)\citenamefont
  {Braunstein}, \citenamefont {Caves},\ and\ \citenamefont
  {Milburn}}]{BRAUNSTEIN1996135}%
  \BibitemOpen
  \bibfield  {author} {\bibinfo {author} {\bibfnamefont {S.~L.}\ \bibnamefont
  {Braunstein}}, \bibinfo {author} {\bibfnamefont {C.~M.}\ \bibnamefont
  {Caves}},\ and\ \bibinfo {author} {\bibfnamefont {G.}~\bibnamefont
  {Milburn}},\ }\href {https://doi.org/https://doi.org/10.1006/aphy.1996.0040}
  {\bibfield  {journal} {\bibinfo  {journal} {Annals of Physics}\ }\textbf
  {\bibinfo {volume} {247}},\ \bibinfo {pages} {135} (\bibinfo {year}
  {1996})}\BibitemShut {NoStop}%
\bibitem [{\citenamefont {Agarwal}\ and\ \citenamefont
  {Davidovich}(2022)}]{agarwal2022quantifying}%
  \BibitemOpen
  \bibfield  {author} {\bibinfo {author} {\bibfnamefont {G.}~\bibnamefont
  {Agarwal}}\ and\ \bibinfo {author} {\bibfnamefont {L.}~\bibnamefont
  {Davidovich}},\ }\href@noop {} {\bibfield  {journal} {\bibinfo  {journal}
  {Physical Review Research}\ }\textbf {\bibinfo {volume} {4}},\ \bibinfo
  {pages} {L012014} (\bibinfo {year} {2022})}\BibitemShut {NoStop}%
\bibitem [{\citenamefont {Demkowicz-Dobrza{\'n}ski}\ \emph
  {et~al.}(2012)\citenamefont {Demkowicz-Dobrza{\'n}ski}, \citenamefont
  {Ko{\l}ody{\'n}ski},\ and\ \citenamefont
  {Gu{\c{t}}{\u{a}}}}]{demkowicz2012elusive}%
  \BibitemOpen
  \bibfield  {author} {\bibinfo {author} {\bibfnamefont {R.}~\bibnamefont
  {Demkowicz-Dobrza{\'n}ski}}, \bibinfo {author} {\bibfnamefont
  {J.}~\bibnamefont {Ko{\l}ody{\'n}ski}},\ and\ \bibinfo {author}
  {\bibfnamefont {M.}~\bibnamefont {Gu{\c{t}}{\u{a}}}},\ }\href@noop {}
  {\bibfield  {journal} {\bibinfo  {journal} {Nature communications}\ }\textbf
  {\bibinfo {volume} {3}},\ \bibinfo {pages} {1} (\bibinfo {year}
  {2012})}\BibitemShut {NoStop}%
\bibitem [{\citenamefont {Zwierz}\ \emph {et~al.}(2012)\citenamefont {Zwierz},
  \citenamefont {P\'erez-Delgado},\ and\ \citenamefont
  {Kok}}]{PhysRevA.85.042112}%
  \BibitemOpen
  \bibfield  {author} {\bibinfo {author} {\bibfnamefont {M.}~\bibnamefont
  {Zwierz}}, \bibinfo {author} {\bibfnamefont {C.~A.}\ \bibnamefont
  {P\'erez-Delgado}},\ and\ \bibinfo {author} {\bibfnamefont {P.}~\bibnamefont
  {Kok}},\ }\href {https://doi.org/10.1103/PhysRevA.85.042112} {\bibfield
  {journal} {\bibinfo  {journal} {Phys. Rev. A}\ }\textbf {\bibinfo {volume}
  {85}},\ \bibinfo {pages} {042112} (\bibinfo {year} {2012})}\BibitemShut
  {NoStop}%
\bibitem [{\citenamefont {Zwierz}\ \emph {et~al.}(2010)\citenamefont {Zwierz},
  \citenamefont {P\'erez-Delgado},\ and\ \citenamefont
  {Kok}}]{PhysRevLett.105.180402}%
  \BibitemOpen
  \bibfield  {author} {\bibinfo {author} {\bibfnamefont {M.}~\bibnamefont
  {Zwierz}}, \bibinfo {author} {\bibfnamefont {C.~A.}\ \bibnamefont
  {P\'erez-Delgado}},\ and\ \bibinfo {author} {\bibfnamefont {P.}~\bibnamefont
  {Kok}},\ }\href {https://doi.org/10.1103/PhysRevLett.105.180402} {\bibfield
  {journal} {\bibinfo  {journal} {Phys. Rev. Lett.}\ }\textbf {\bibinfo
  {volume} {105}},\ \bibinfo {pages} {180402} (\bibinfo {year}
  {2010})}\BibitemShut {NoStop}%
\bibitem [{\citenamefont {Lahiri}\ \emph {et~al.}(2018)\citenamefont {Lahiri},
  \citenamefont {Gharavi}, \citenamefont {Baugh},\ and\ \citenamefont
  {Muralidharan}}]{PhysRevB.98.125417}%
  \BibitemOpen
  \bibfield  {author} {\bibinfo {author} {\bibfnamefont {A.}~\bibnamefont
  {Lahiri}}, \bibinfo {author} {\bibfnamefont {K.}~\bibnamefont {Gharavi}},
  \bibinfo {author} {\bibfnamefont {J.}~\bibnamefont {Baugh}},\ and\ \bibinfo
  {author} {\bibfnamefont {B.}~\bibnamefont {Muralidharan}},\ }\href
  {https://doi.org/10.1103/PhysRevB.98.125417} {\bibfield  {journal} {\bibinfo
  {journal} {Phys. Rev. B}\ }\textbf {\bibinfo {volume} {98}},\ \bibinfo
  {pages} {125417} (\bibinfo {year} {2018})}\BibitemShut {NoStop}%
\bibitem [{\citenamefont {Aharonov}\ and\ \citenamefont
  {Vaidman}(2008)}]{Aharonov2008}%
  \BibitemOpen
  \bibfield  {author} {\bibinfo {author} {\bibfnamefont {Y.}~\bibnamefont
  {Aharonov}}\ and\ \bibinfo {author} {\bibfnamefont {L.}~\bibnamefont
  {Vaidman}},\ }\bibinfo {title} {The two-state vector formalism: An updated
  review},\ in\ \href {https://doi.org/10.1007/978-3-540-73473-4_13} {\emph
  {\bibinfo {booktitle} {Time in Quantum Mechanics}}},\ \bibinfo {editor}
  {edited by\ \bibinfo {editor} {\bibfnamefont {J.}~\bibnamefont {Muga}},
  \bibinfo {editor} {\bibfnamefont {R.~S.}\ \bibnamefont {Mayato}},\ and\
  \bibinfo {editor} {\bibfnamefont {{\'I}.}~\bibnamefont {Egusquiza}}}\
  (\bibinfo  {publisher} {Springer Berlin Heidelberg},\ \bibinfo {address}
  {Berlin, Heidelberg},\ \bibinfo {year} {2008})\ pp.\ \bibinfo {pages}
  {399--447}\BibitemShut {NoStop}%
\end{thebibliography}%
\end{document}